\documentclass{article}

\usepackage{amsmath}
\usepackage{amssymb}
\usepackage{amsfonts}
\usepackage{bm}
\usepackage[dvips,xdvi]{graphicx}

\begin{document}


%
\title{\bf
Self-Consistent Field Model Simulations for
Statistics of Amorphous Polymer Chains in Crystalline Lamellar Structures}

\author{Takashi Uneyama\textsuperscript{1}\footnote{
E-mail: uneyama@se.kanazawa-u.ac.jp,
Tel: +81-76-264-6221, Fax: +81-76-234-4829}, Takafumi Miyata\textsuperscript{2},
and Koh-hei Nitta\textsuperscript{1} \\
\\
\textsuperscript{1} Institute of Science and Engineering, Kanazawa
University, \\
Kakuma, Kanazawa 920-1192, Japan \\
\textsuperscript{2} Graduate School of Engineering, Nagoya University, \\
Chikusa, Nagoya 464-8603, Japan
}
\date{}

\maketitle



\begin{abstract}
We calculate statistical properties of amorphous polymer chains between
crystalline lamellae by self-consistent field model
 simulations. In our model, an amorphous subchain is modelled as a
 polymer chain of which ends are
 grafted onto the crystal-amorphous interfaces. The crystal-amorphous
 interfaces are expressed as impenetrable surfaces. We incorporate
 the interaction between segments to satisfy the incompressible
 condition for the segment density field.
 The simulation results show that amorphous polymer chains feel thin
 potential layers, which are mainly repulsive, near the crystal-amorphous
 interfaces. The impenetrable and incompressible conditions affect the statistics of polymer
 chains and the chain statistics becomes qualitatively different from
 the ideal Gaussian chain statistics without any constraints. We show
 the effects of the system size and the graft density to statistical
 quantities. We also show that the tie subchain statistics obey rather
 simple statistics.
\end{abstract}


\maketitle


\section{Introduction}
\label{introduction}

Crystalline polymers form various superstructures such as the crystalline lamellar
structure, which consists of crystal and amorphous
layers, and the spherulite structure\cite{Mandelkern-chapter,Strobl-book,Seguela-2005,Young-Lovell-book}.
The mechanical properties of crystalline polymers depend on these
superstructures.
The crystalline lamellae strongly affect mechanical properties, and the relation between the crystalline lamellar structure and
mechanical properties have been studied extensively\cite{Strobl-book,Seguela-2005}.
Especially, the statistics of tie subchains and tie molecules in crystalline
lamellae is reported to be important for the
mechanical properties\cite{Krigbaum-Roe-Smith-1964,Weiner-Berman-1984,Boyd-1985,Lu-Qian-Brown-1995,Takayanagi-Nitta-1997,Nitta-Takayanagi-1999,Nitta-Takayanagi-2003}.
Although considerable studies have been made
for the statistics of tie subchains and tie molecules, it is not yet fully
understood.
Various theoretical models have
been proposed for loop and tie subchains, and for tie molecules\cite{Flory-1962,DiMarzio-Guttman-Hoffman-1980,Guttman-DiMarzio-1982,Leermakers-Scheutjens-Gaylord-1984,Flory-Yoon-Dill-1984,Marqusee-Dill-1986,Lin-Argon-1994,Huang-Borwn-1999}. For example,
so-called the gambler's ruin model\cite{Guttman-DiMarzio-1982} gives the statistics of amorphous
subchains between crystalline lamellae. The Huang-Brown model\cite{Huang-Borwn-1999} gives the
estimate for the tie molecule fraction based on the ideal Gaussian chain
statistics in melt before the crystallization.
{In what follows, we may call the Gaussian chain
statistics in absence of inhomogeneous external fields as ``the ideal
Gaussian chain statistics.''}

Simulations have been also utilized to study the statistics of tie
subchains and tie
molecules\cite{Mathur-Mattice-1987,Mansfield-1990,Balijepalli-Rutledge-1998,Balijepalli-Rutledge-1998a,Balijepalli-Rutledge-2000}.
For example, we can compute
various statistical quantities for tie subchains by the lattice Monte Carlo
simulations and validate theoretical predictions. 
Due to the limitation of the computational resources,
in most cases we need to employ
coarse-grained models rather than atomistic models. These coarse-grained
models involve some approximations and simplifications, not all of which
are justified. To simplify the model,
in some cases (especially for off-lattice systems), the ideal Gaussian
chain statistics is assumed without justification.
However, the polymer chains in the amorphous phase interact each other
rather strongly. The ideal Gaussian chain statistics is reproduced as a
result of the screening effect\cite{deGennes-book}, and naively, we expect that such a
screening can be realized only in a bulk system. {If the
screening is not perfect, or if an external field is applied, the
statistics of a polymer chain deviates from the ideal Gaussian chain
statistics. In such cases, we need to take into account the contribution of
the effective potential field.}

In microphase separation structures of block copolymer melts, statistics
of polymer
chains are known to be qualitatively different from one of the bulk
systems. For example,
a diblock copolymer chain in a strongly segregated lamellar microphase separation structure is
highly stretched, and the chain size (which is comparable to the
lamellar period) scales as $N^{2/3}$ (with $N$ being the polymerization
index)\cite{Ohta-Kawasaki-1986,Kawakatsu-book,Matsen-chapter}. This is
qualitatively different from the $N$-dependence of
an ideal Gaussian chain size in a bulk, $N^{1/2}$. Polymer chains in the melt brush (dry
brush) systems do not behave as ideal Gaussian chains, neither\cite{Milner-Witten-Cates-1988,Kawakatsu-book,Matsen-chapter}.
A polymer chain in a melt brush feels an inhomogeneous effective potential
(pressure) field, and the free chain ends segregate into the surface of
the brush. In both cases, the interaction between segments (or the
incompressible condition for the segment density field) plays an
important role. This implies that the interaction between segments can be
also important in the crystalline lamellar systems.

Recently, Milner\cite{Milner-2011} investigated the nucleation process of
polyethylene by simulations.
To estimate the free energy barrier for the nucleation, he calculated
the interfacial free energy between the bulk amorphous and the nucleus.
The interface between the amorphous and the nucleus was modelled as the
interface between infinitely large amorphous and crystal phases.
Milner introduced the mean field potential field which
makes the segment density spatially homogeneous
(to satisfy the incompressible condition), and performed the
lattice self-consistent field (SCF)
simulations\cite{Kawakatsu-book,Matsen-chapter,Muller-Schmid-2005}.
The simulation results
showed the existence of a thin repulsive layer near the interface. Interestingly,
the contributions of the mean field potential is not large except this
repulsive layer. Although the repulsive layer is thin (just one lattice layer
in the lattice SCF model), it affects the statistics of the polymer chains in
the amorphous region rather strongly.

We consider that the situation is similar for the amorphous layers between
crystalline lamellae. Namely, we expect that
repulsive layers exist near the crystal-amorphous interfaces, and the
statistics of the polymer chains in the amorphous region is largely
affected by these repulsive layers. In this work, we propose a SCF model
for the polymer subchains in the amorphous region between crystalline
lamellae. We perform SCF simulations and show that the repulsive
potential layers exist near the crystal-amorphous interfaces. Based on
the simulation results, we calculate some statistical quantities of the
amorphous subchains and discuss the statistics of the tie subchain in detail.

\section{Model}
\label{model}

\subsection{Self-Consistent Field Model}
\label{self_consistent_field_model}

\begin{figure}[h!]
 \centering
 {\includegraphics[width=0.95\linewidth,clip]{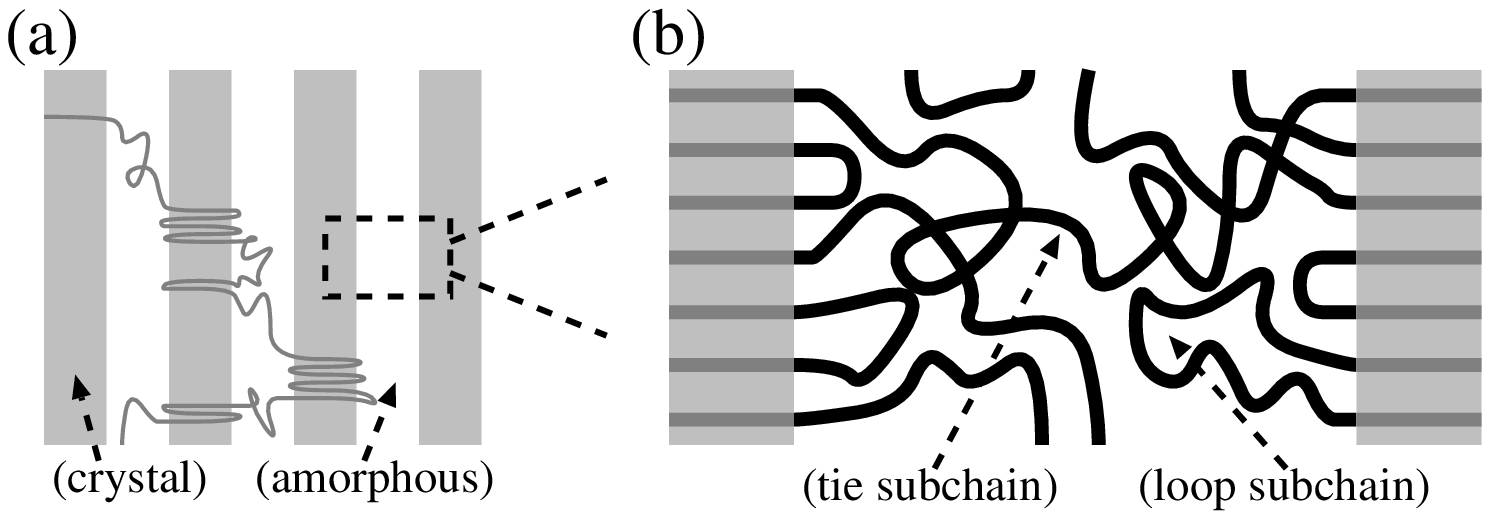}}
 \caption{}
 \label{lamellar_image}
\end{figure}

We illustrate the schematic images of the crystalline lamellar structure
and an amorphous region between crystalline lamellae in Figure
\ref{lamellar_image} (a) and (b).
In this work, we model an amorphous subchain between the
crystalline lamellae as a grafted chain\cite{Milner-2011}. Both ends of a subchain are
grafted onto the crystal-amorphous interfaces.
For simplicity, we assume that the polymer chains are
sufficiently long and the effects of the chain ends are negligible.
(In other words, we do not consider the cilia subchains.)
The interaction between segments is expressed via the
mean field potential, which is determined self-consistently.
From the symmetry, the one-dimensional system is sufficient for our simulations.
We express the
positions of the crystal-amorphous interfaces as $x = 0$ and $x =
L$. ($L$ represents the thickness of the amorphous layer. In the
followings we call $L$ as the system size.)

The crystalline lamellar structure is formed when the melt is cooled.
As known well, the thermal histroy, such as the quenching and
annealing conditions, strongly affects the crystalline structure and
subchain statistics\cite{Mandelkern-chapter,Strobl-book,Seguela-2005,Young-Lovell-book}.
In this work, we consider the systems after sufficiently long annealing,
and assume that the subchains in the system are well equilibrated. Under
such an assumption, the subchains obey the (local) equilibrium statistics, and
the standard SCF model can be reasonably utilized.

The application of the SCF model to systems with hard and impenetrable walls have been
done in various works\cite{DiMarzio-1965,Dolan-Edwards-1974,deGennes-book}.
Following the standard SCF model, the statistics of amorphous subchains can be
described by using the path integral field.
(See Appendix \ref{derivation_of_the_self_consistent_field_equations}
for the detailed derivation of the SCF model.)
As we mentioned, the system can be considered as one-dimensional. Thus,
from the symmetry, the path integral field does not depend
on $y$ nor $z$. We express the path integral field (and other
fields) as a function of $x$.
The path integral field $q(x,s)$ obeys the Edwards equation
\begin{equation}
 \label{edwards_equation}
 \frac{\partial q(x,s)}{\partial s}
 = \frac{b^{2}}{6} \frac{\partial q(x,s)}{\partial x^{2}} - v(x)
 q(x,s)
\end{equation}
where $s$ is the segment index, $b$ is the segment size, and $v(x)$ is
the mean field potential.
The boundary and initial conditions for eq \eqref{edwards_equation} are
\begin{equation}
 \label{boundary_condition_edwards_equation}
  q(0,s) = q(L,s) = 0 ,
\end{equation}
\begin{equation}
 \label{initial_condition_edwards_equation}
  q(x,0) = b [ \delta(x - \epsilon) + \delta(x - L + \epsilon) ] .
\end{equation}
Here $\epsilon$ is a small positive constant, and we take the limit of
$\epsilon \to 0$ at the end of the calculation.
Eq \eqref{boundary_condition_edwards_equation} is the absorbing boundary
condition which represents the effect of impenetrable regions\cite{DiMarzio-1965}.
In eq \eqref{initial_condition_edwards_equation}, {we have introduced the
numerical factor $b$ so that 
the path integral field $q(x,s)$ becomes dimensionless. (This numerical
factor does not affect any thermodynamic properties such as the density.)}

The density field $\rho(x)$ is calculated from the path integral field
$q(x,s)$ as
\begin{equation}
 \label{density_from_path_integral}
 \rho(x) = \frac{\bar{\sigma}}{\mathcal{Q}}
  \sum_{n = 1}^{\infty} e^{\mu n / k_{B} T} \int_{0}^{n} ds \, q(x,s) q(x,n - s)
\end{equation}
where $\bar{\sigma}$ is the graft density (or the injection density),
$\mathcal{Q}$ is the single
chain partition function (normalized by the interfacial area), and $\mu$
is the chemical potential for 
segments. (This chemical potential is required to control the total number
of segment in the system. {In our model, $\mu$ works as the
Lagrange multiplier, and is automatically and uniquely determined, as we show below.})
The single chain partition function is defined as follows:
\begin{equation}
 \label{single_chain_partition_function}
  \mathcal{Q} = 
  \sum_{n = 1}^{\infty} e^{\mu n / k_{B} T} \int_{0}^{L} dx \, q(x,0)
  q(x,n) .
\end{equation}
We determine the chemical potential $\mu$ so that the spatial average
of the density field becomes a given value $\bar{\rho}$ as
\begin{equation}
 \label{chemical_potential_condition}
  \bar{\rho} = \frac{1}{L} \int_{0}^{L} dx \, \rho(x) .
\end{equation}

Finally, the mean field potential $v(x)$ is determined from the density
field $\rho(x)$. In this work, we employ the harmonic potential
\begin{equation}
 \label{potential_from_density}
 v(x) = \frac{1}{k_{B} T \bar{\rho}^{2} \kappa} [\rho(x) - \bar{\rho}]
\end{equation}
where $\kappa$ is the compressibility\cite{Kawakatsu-book}. We assume
that the segment density field is
almost constant ($\rho(x) \approx \bar{\rho}$) due to the incompressible
condition, and thus $\kappa$
should be sufficiently small.
The equilibrium solution is given as the set of fields which satisfies eqs
\eqref{edwards_equation}-\eqref{potential_from_density}
self-consistently.

{
The free energy of the system can be calculated from the self-consitent
set of fields. For the harmonic
potential (eq \eqref{potential_from_density}), the explicit expression of the free energy becomes simply as follows:
\begin{equation}
 \label{free_energy}
   \frac{\mathcal{F}}{k_{B} T}
   = - \frac{1}{2 k_{B} T \bar{\rho}^{2} \kappa}
   \int_{0}^{L} dx \, [\rho(x) - \bar{\rho}]^{2}
   - \bar{\sigma} \ln \frac{e \mathcal{Q}}{\bar{\sigma}}
\end{equation}
The free energy per unit volume is useful to
study the thermodynamic stability of the obtained structure. In our
model, the free energy per unit volume is simply expressed as $f \equiv
\mathcal{F} / L$. To obtain the thermodynamic equilibrium structures, we
need to minimize the free energy per unit volume with respect to the system
size. Such a procedure is essential for microphase separation structures of
block copolymers. However, the cyrstalline lamellar structure is not
true thermodynamic equilibrium structure and thus we expect that there
is no optimal system size which minimizes the free energy per unit
volume. Therefore, in this work we do not minimize the free energy per
unit volume with respect to the system size.
}

\subsection{Numerical Scheme}
\label{numerical_scheme}

In this subsection, we describe the discretization method and numerical
scheme for the SCF simulations. We use the dimensionless units for
simulations, by setting $b = 1$ and $k_{B} T = 1$. 
There are roughly two numerical methods to solve the SCF model.
One is the Fourier spectral method\cite{Matsen-chapter} and another is the real space method\cite{Fredrickson-Ganesan-Drolet-2002,Muller-Schmid-2005}.
Generally, the Fourier spectral method is preferred if the symmetry of the
system is known. This is because the Fourier spectral method gives accurate results
with relatively small numerical costs.
However, in our case, the Fourier spectral method does not work as usual,
due to the singularity of the mean field potential field.
(We will discuss this point later.) Thus we employ the real space method, and
discretize $q(x,s)$, $\rho(x)$ and $v(x)$ by using a one dimensional regular mesh.
We divide the system into $m$ mesh points, and express the position of the $i$-th
mesh point as $x_{i} = (i + 1 / 2) L / m$ $(i = 0,1,2,\dots,m - 1)$.
For convenience, we express the mesh size as $h \equiv L /
m$. Also, we express the discretized fields as $q_{i}(s)$,
$\rho_{i}$, and $v_{i}$ in the followings.

We approximate the spatial derivative in eq \eqref{edwards_equation} by the
central difference scheme. Then eq \eqref{edwards_equation} with
the boundary condition \eqref{boundary_condition_edwards_equation} can be
discretized as
\begin{equation}
 \label{edwards_equation_discretized}
 \frac{dq_{i}(s)}{ds}
  = - \sum_{j} C_{ij} q_{j}(s)
\end{equation}
with the coefficient matrix $C_{ij}$ defined as
\begin{equation}
 \label{coefficient_matrix}
 C_{ij} \equiv
  \begin{cases}
   \displaystyle 1 / 3 h^{2} + v_{i} & (i = j), \\
   \displaystyle - 1 / 6 h^{2} & (i = j \pm 1), \\
   0 & (\text{otherwise}).
  \end{cases}
\end{equation}
The initial condition \eqref{initial_condition_edwards_equation} can be
discretized as follows:
\begin{equation}
 \label{initial_condition_edwards_equation_discretized}
 q_{i}(0) = \frac{1}{h} (\delta_{i,0} + \delta_{i,m - 1}) .
\end{equation}

We can solve eq \eqref{edwards_equation_discretized}, by
using the eigenvalues and eigenvectors of $C_{ij}$ (just like the
Fourier spectral method \cite{Matsen-chapter}),
\begin{equation}
 \label{edwards_equation_discretized_integrated}
  q_{i}(s)
  = \sum_{j} U_{ij} e^{- s \lambda_{j}} p_{j}
\end{equation}
where $\lambda_{j}$ and $U_{ij}$ represent the $j$-th eigenvalue and
eigenvector, respectively. $p_{i}$ is defined as
\begin{equation}
 \label{initial_vector}
  p_{i} \equiv \sum_{j} U_{ji} q_{j}(0) = \frac{1}{h} (U_{0,i} + U_{m - 1,i}).
\end{equation}
We can rewrite the discretized versions of
eqs \eqref{density_from_path_integral}-\eqref{potential_from_density}
by using the eigenvalues and eigenvectors.
Eqs \eqref{density_from_path_integral} and
\eqref{single_chain_partition_function} become
\begin{equation}
 \label{discretized_density_from_path_integral}
   \rho_{i} = \frac{\bar{\sigma}}{\mathcal{Q}} G_{i}
\end{equation}
and
\begin{equation}
 \label{discretized_single_chain_partition_function}
  \mathcal{Q}
   = h \sum_{i} \frac{e^{- \lambda_{i} + \mu}}{1 - e^{- \lambda_{i} + \mu}} p_{i}^{2}
\end{equation}
with $G_{i}$ being defined as
\begin{equation}
 \label{weight_vector_definition}
  G_{i}
  \equiv \sum_{j,k} U_{ij} U_{ik}
  \frac{1}{\lambda_{k} - \lambda_{j}}
  \left[ \frac{e^{-\lambda_{j} + \mu}}{1 - e^{- \lambda_{j} + \mu}}
  - \frac{e^{- \lambda_{k} + \mu}}{1 - e^{- \lambda_{k} + \mu}}
  \right]
  p_{j} p_{k} .
\end{equation}
(In the case of $j = k$ or degenerated eigenvalues, the factor $1 /
(\lambda_{k} - \lambda_{j})$ in eq \eqref{weight_vector_definition}
diverges. To avoid this divergence, for such cases we expand the
expression in the summation into the series of $\lambda_{k} -
\lambda_{j}$.)
Eq \eqref{chemical_potential_condition} is discretized as
\begin{equation}
 \label{chemical_potential_condition_modified}
  \bar{\rho} 
  = \frac{h \bar{\sigma}}{\mathcal{Q} L} \sum_{i}
  \frac{e^{-\lambda_{i} +\mu}}
  {[1 - e^{-\lambda_{i} +\mu}]^{2}}p_{i}^{2} .
\end{equation}
The chemical potential $\mu$ in eqs
\eqref{discretized_single_chain_partition_function} and
\eqref{weight_vector_definition}
is numerically determined to satisfy eq \eqref{chemical_potential_condition_modified}.
Eq \eqref{potential_from_density} is simply discretized as
\begin{equation}
 \label{potential_from_density_discretized}
 v_{i} = \frac{1}{\bar{\rho}^{2} \kappa} (\rho_{i} - \bar{\rho}) .
\end{equation}

We perform simulations by solving eqs
\eqref{edwards_equation_discretized}-\eqref{potential_from_density_discretized}
numerically.
We employ the iteration method, which is widely employed in the SCF simulations\cite{Kawakatsu-book,Muller-Schmid-2005}.
The numerical scheme to solve the set of discretized equations is as
follows.
\begin{enumerate}
 \item Set an initial guess for the potential field $v_{i}$ as $v_{i} = 0$.
 \item \label{scf_step_diagonalization}
       Calculate the coefficient matrix $C_{ij}$ (eq \eqref{coefficient_matrix})
       from $v_{i}$ and calculate the eigenvalues and eigenvectors,
       $\lambda_{j}$ and $U_{ij}$.
 \item Calculate the chemical potential $\mu$ by solving eq
       \eqref{chemical_potential_condition_modified} numerically.
       (In this work we employ the bisection method.)
 \item Calculate the density field $\rho_{i}$ from the eigenvalues $\lambda_{j}$,
       eigenvectors $U_{ij}$, and chemical potential
       $\mu$ by eqs
       \eqref{initial_vector}-\eqref{weight_vector_definition}.
 \item \label{scf_step_potential_update}
       Update the potential field $v_{i}$ to $v_{i}^{\text{(new)}}$ as
       \begin{equation}
	v_{i}^{\text{(new)}} = (1 - \theta) v_{i} + \frac{\theta}{\bar{\rho}^{2}\kappa} (\rho_{i} - \bar{\rho})
       \end{equation}
       where $\theta$ is a small positive constant (the update parameter).
 \item \label{scf_step_residual_calculation}
       Evaluate the residue $R$ for $v_{i}$:
       \begin{equation}
	R \equiv \frac{1}{m} \sum_{i} \left| v_{i} - \frac{1}{\bar{\rho}^{2}\kappa }
				       (\rho_{i} - \bar{\rho}) \right| .
       \end{equation}
       If the residue $R$ is larger than a threshold value, then go to Step
       \ref{scf_step_diagonalization}.
\end{enumerate}
For the calculation of the eigenvalues and eigenvectors at Step \ref{scf_step_diagonalization}, we
utilize the DSTEV subroutine in LAPACK (the QR method) \cite{lapack-usersguide}.
The eigenvalues and eigenvectors can be successfully calculated within
an acceptable numerical error range.
(We have also examined the DSTEMR subroutine, which utilizes the MRRR method\cite{Dhillon-Parlett-Vomel-2006}, and
have observed that the obtained eigenvalues and eigenvectors are
almost the same as ones obtained by the DSTEV subroutine.)
After the calculation converges, physical quantities can be
calculated by using quantities such as
$\lambda_{i}$ and $U_{ij}$. For example, the fraction of the subchain
which consists of $n$ segments (the segment number distribution) is calculated as
\begin{equation}
 \phi(n) = \frac{h}{\mathcal{Q}} \sum_{i} e^{-n (\lambda_{i} - \mu)}
  p_{i}^{2} .
\end{equation}
{
By discretizing eq \eqref{free_energy}, we have
the following expression for the free energy per unit volume:
\begin{equation}
 f = - \frac{1}{2 \bar{\rho}^{2} \kappa m} \sum_{i} (\rho_{i} - \bar{\rho})^{2}
  - \frac{\bar{\sigma}}{L}
  \left( \ln \frac{\mathcal{Q}}{\bar{\sigma}} + 1 \right) .
\end{equation}
}

The statistics of the tie and loop subchains can be calculated by changing
the initial condition (the graft condition). For the calculation of the loop subchain of which
ends are grafted to the interface at $x = 0$, we use the initial condition
\begin{equation}
 \label{initial_condition_edwards_equation_loop_discretized}
 q_{i}(0) = \frac{1}{h} \delta_{i,0}
\end{equation}
instead of eq \eqref{initial_condition_edwards_equation_discretized}.
The calculation with the initial condition eq
\eqref{initial_condition_edwards_equation_loop_discretized} can be
easily performed by replacing the vector $p_{i}$ by the following vector:
\begin{equation}
 \tilde{p}_{i} \equiv \frac{1}{h} U_{0,i} .
\end{equation}
The (discretized) density field of the loop subchains of which ends are grafted to the
interface at $x = 0$ is 
\begin{equation}
 \rho_{l,i} = \frac{\bar{\sigma}}{\mathcal{Q}}
  \tilde{G}_{i}
\end{equation}
where $\tilde{G}_{i}$ is defined by eq \eqref{weight_vector_definition}
with $G_{i}$ and $p_{i}$ replaced by $\tilde{G}_{i}$ and
$\tilde{p}_{i}$, respectively. From the symmetry, the density
field of the loop subchains of which ends are grafted to the interface at $x =
L$ is simply given as
\begin{equation}
 \rho_{l,i}' = \rho_{l,m - 1 - i}
\end{equation}
and the density field of tie subchains is calculated as
\begin{equation}
 \rho_{t,i} = \rho_{i} - \rho_{l,i} - \rho_{l,i}' .
\end{equation}
The segment number distributions for the tie and loop subchains are calculated in a
similar way.
\begin{align}
 & \phi_{l}(n) = \frac{2 h}{\mathcal{Q}} \sum_{i} e^{- n (\lambda_{i} -
 \mu)} \tilde{p}_{i}^{2} , \\
 & \phi_{t}(n) = \phi(n) - \phi_{l}(n) .
\end{align}
For convenience, in the followings, we call $\phi_{l}(n)$ and
$\phi_{t}(n)$ as the tie and loop subchain distribution functions, respectively.

\section{Results}
\label{results}

We perform the SCF simulations with the numerical scheme shown in
Section \ref{numerical_scheme}.
We fix the average
segment density as $\bar{\rho} = 1$, and vary the graft density
$\bar{\sigma}$ and the system size $L$.
We also fix the mesh size as $h = 1 / 8$ and thus the number of
mesh points is $m = 8 L$. The compressibility is
empirically chosen to be $\bar{\kappa} = 0.001$. This value is
sufficiently small to realize almost constant  segment density fields
and approximately satisfy the incompressible condition.
The update parameter $\theta$ is also empirically chosen to make the
iterations stable. Although the optimal value of $\theta$ depends on
other parameters, the typical value is of the order of $10^{-4}$. We set the
tolerance value for the iteration as $10^{-7}$. (The effects of these
parameters are rather minor. If we perform simulations with the
different values of $\theta$ and the tolerance, the obtained results are
almost the same as the original result.)

\subsection{Density and Potential Fields}
\label{density_and_potential_fields}

\begin{figure}[h!]
 \centering
 {\includegraphics[width=0.95\linewidth,clip]{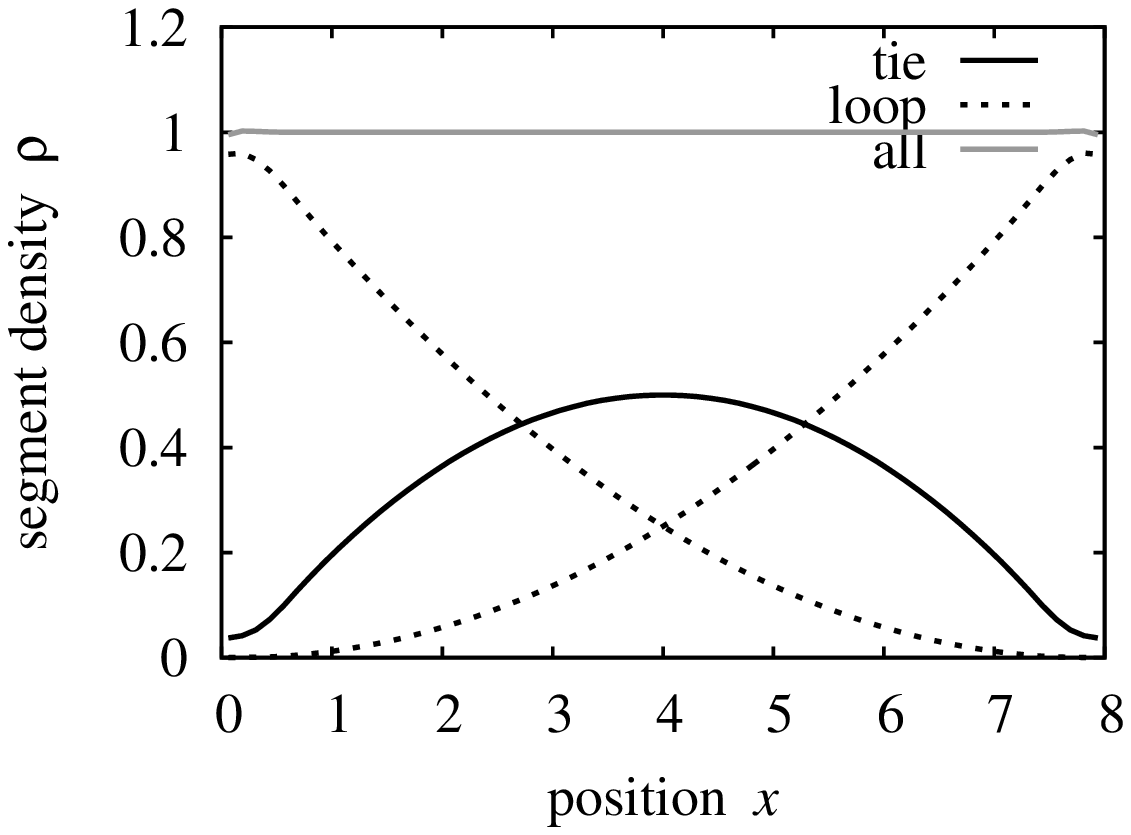}}
 \caption{}
 \label{density_profiles_s1_l8}
\end{figure}

Firstly, we show the profiles of the segment densities of subchains and the
effective potential for segments. Figure \ref{density_profiles_s1_l8} shows the segment density
profiles for the tie, loop, and all subchains for $L = 8$ and $\bar{\sigma}
= 1$, as an example. The segment density of
all subchains is almost constant, which means that the
incompressible condition is approximately satisfied. The loop subchain density field
is high near the crystal-amorphous interface and decrease as the
distance from the interface increases. On the other hand, the tie
subchain density field is very low near the crystal-amorphous
interface and is high near the center of the amorphous region. In other
words, the tie subchain segments are concentrated in 
the center of the amorphous region.

\begin{figure}[c!]
 \centering
 {\includegraphics[width=0.95\linewidth,clip]{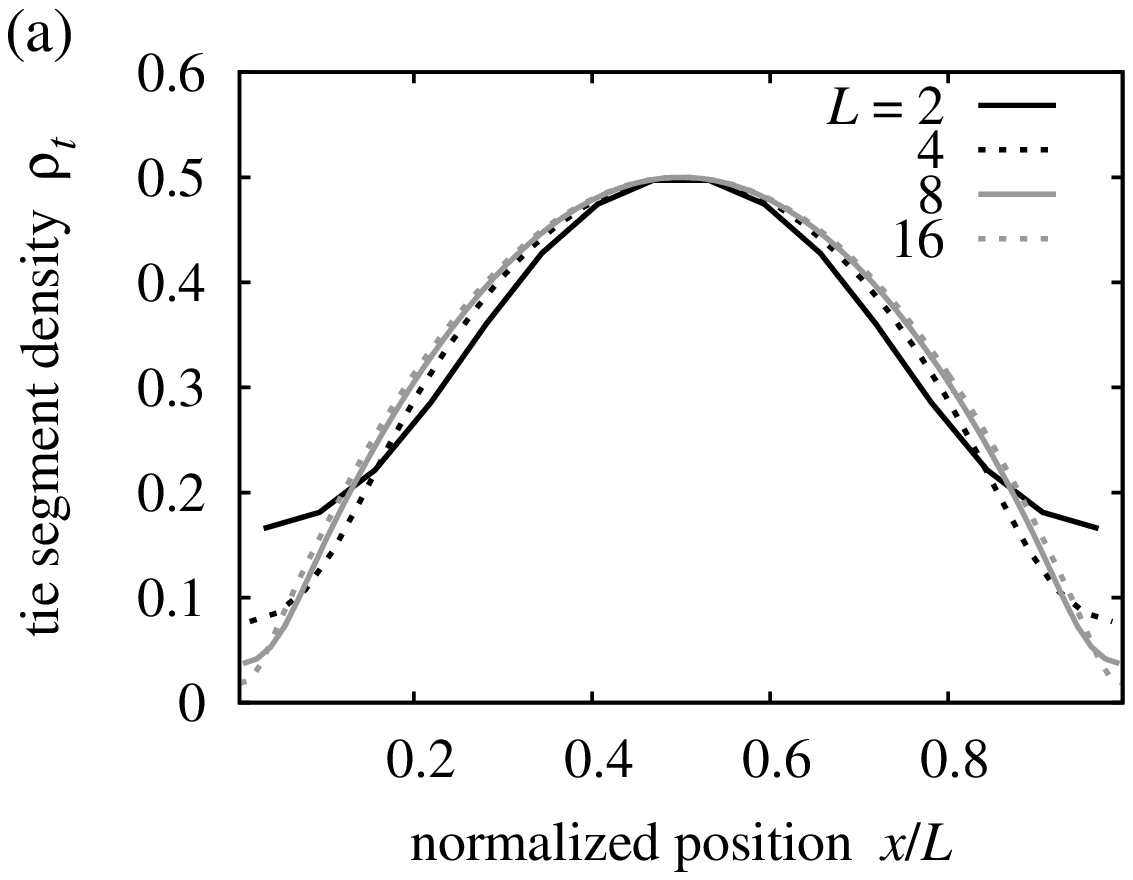}}
 {\includegraphics[width=0.95\linewidth,clip]{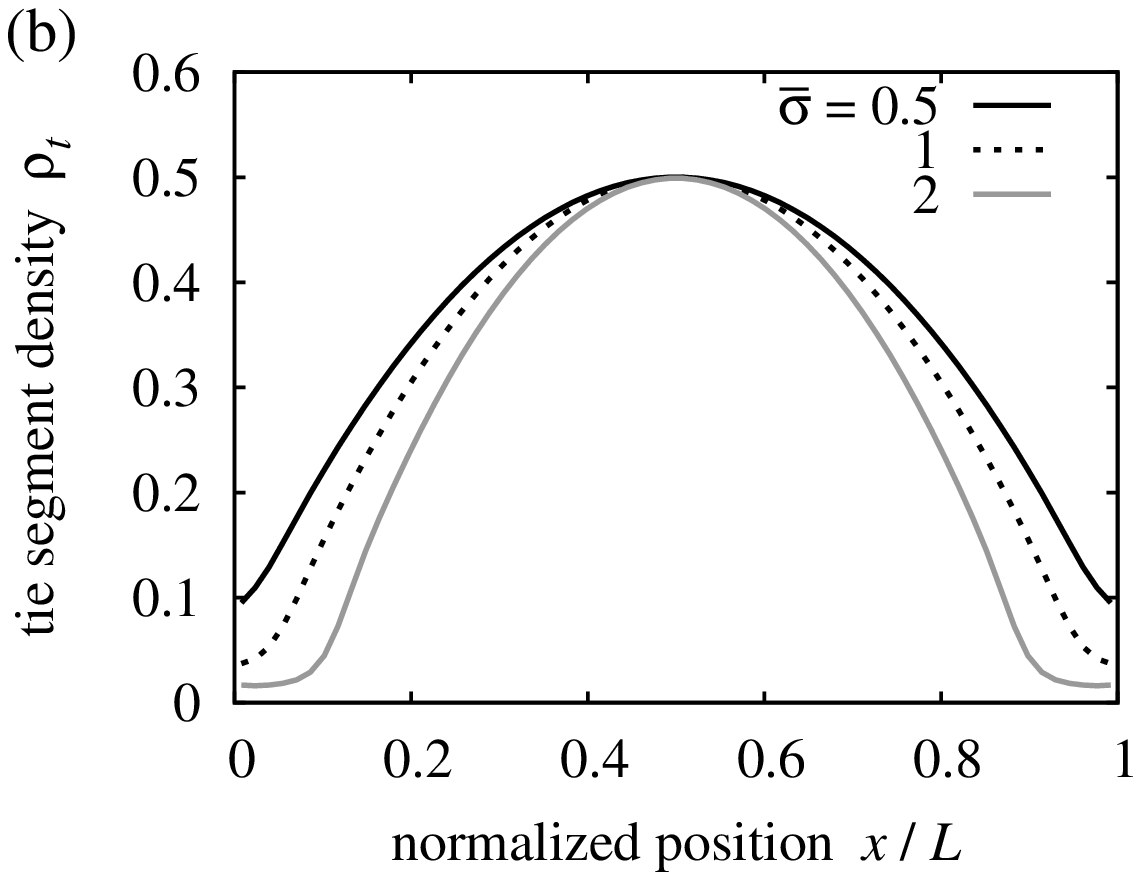}}
 \caption{}
 \label{density_profiles_normalized}
\end{figure}

Interestingly, the tie subchain density seems to be the same
as the total loop subchain densities at the center of the system,
$\rho_{t}(L / 2) = \rho_{l}(L / 2) + \rho_{l}'(L /
2)$. This implies that the relation $\rho_{t}(L / 2) =
\bar{\rho} / 2$ holds.
This can be intuitively understood from the symmetry argument, as follows.
If we find a subchain of one of which segments is at the center of the amorphous region,
the probabilities that one end of the subchain is connected to the
crystal-amorphous interface at $x = 0$ and $x = L$ are the same.
Thus we find the tie subchain and the loop subchain by the equal
probabilities.
Figure \ref{density_profiles_normalized} shows the system size and graft density
dependence of the tie segment density profile. We can observe that $\rho_{t}(L / 2) =
\bar{\rho} / 2$ actually holds for various values of $L$ and $\bar{\sigma}$. The
tie segment density profile near the interface strongly depends on $L$
and $\bar{\sigma}$. The tie segment density near the interface decreases
as $L$ increases or as $\bar{\sigma}$ increases.

\begin{figure}[c!]
 \centering
 {\includegraphics[width=0.95\linewidth,clip]{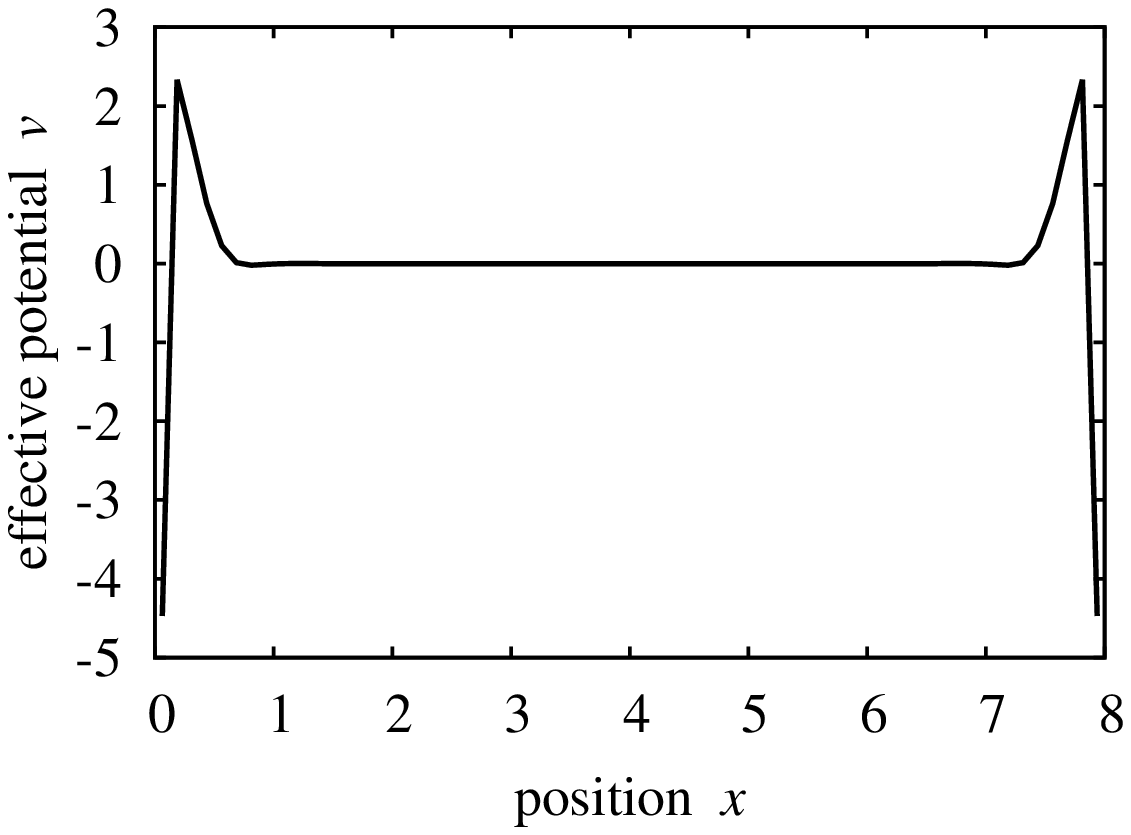}}
 \caption{}
 \label{potential_profile_s1_l8}
\end{figure}

\begin{figure}[c!]
 \centering
 {\includegraphics[width=0.95\linewidth,clip]{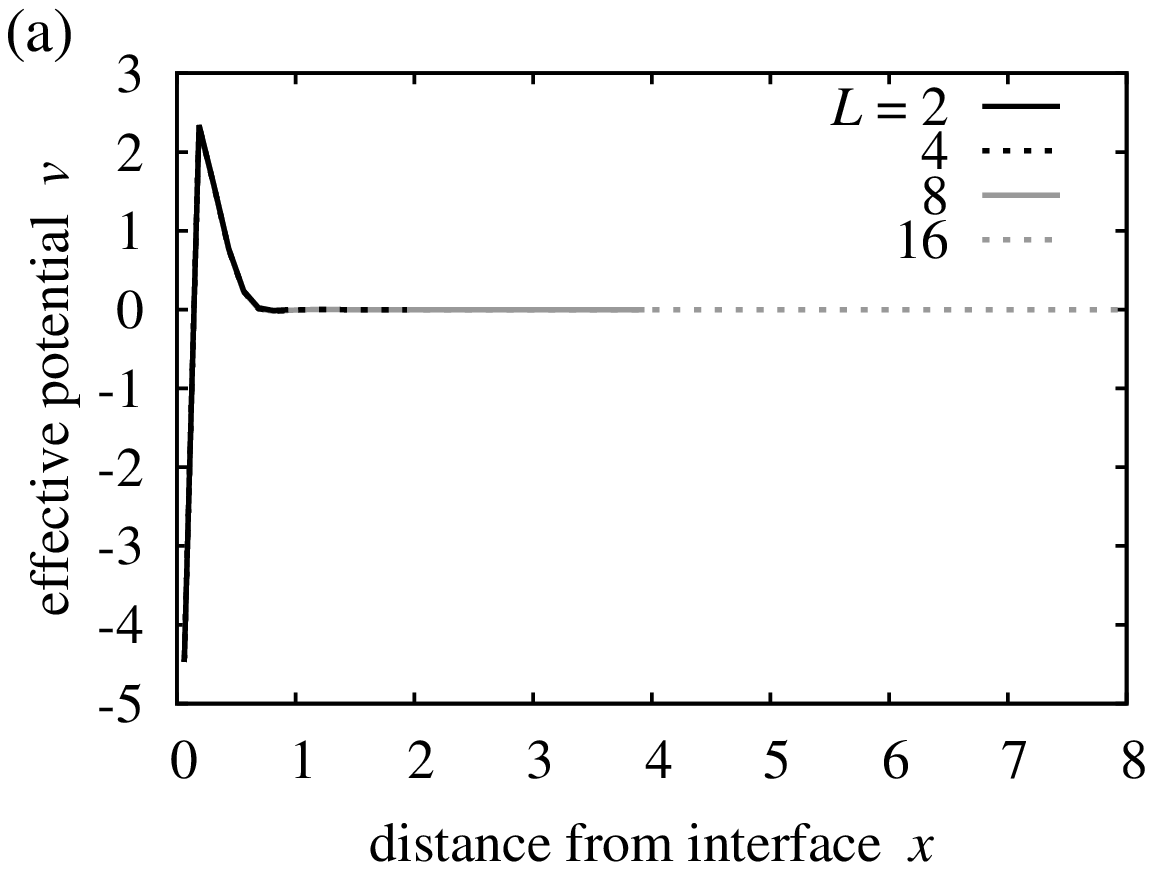}}
 {\includegraphics[width=0.95\linewidth,clip]{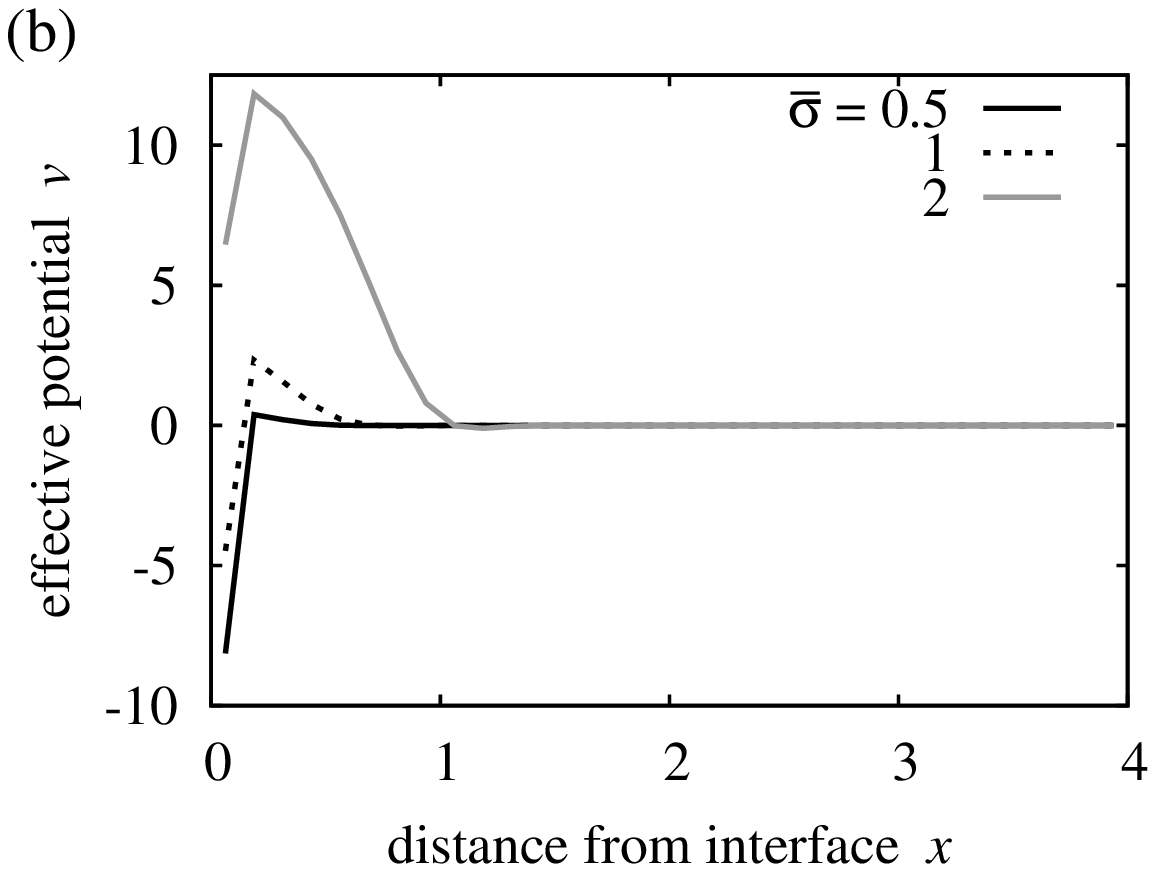}}
 \caption{}
 \label{potential_profile_half}
\end{figure}

Figure \ref{potential_profile_s1_l8} shows the effective mean field
potential profile for segments, for $L = 8$ and $\bar{\sigma}
= 1$ (which are the same parameters as the simulation shown in Figure \ref{density_profiles_s1_l8}).
We can observe that the potential field is almost constant except the
narrow regions near the crystal-amorphous interfaces.
This result is similar to the Milner's simulation results. However,
unlike Milner's results, our result shows that the potential layers near
the interfaces are not purely repulsive.
The potential layer consists of a thin
attractive part and a broad repulsive part. We will discuss this in detail, later.
We show the system size and graft density dependence of the potential
profiles in Figure \ref{potential_profile_half}.
As shown in Figure \ref{potential_profile_half}(a), the potential
profile near the interface is almost independent of the system size $L$.
On the other hand, the graft density $\bar{\sigma}$ affects the
potential profile rather strongly. Thus we expect
that the mean field potential is primarily controlled by the graft density.
The potential barrier near the crystal-amorphous interfaces repel
segments so that the segment density becomes spatially homogeneous. The
required potential barrier to satisfy the incompressible condition
increases as the graft density increases.

\begin{figure}[c!]
 \centering
 {\includegraphics[width=0.95\linewidth,clip]{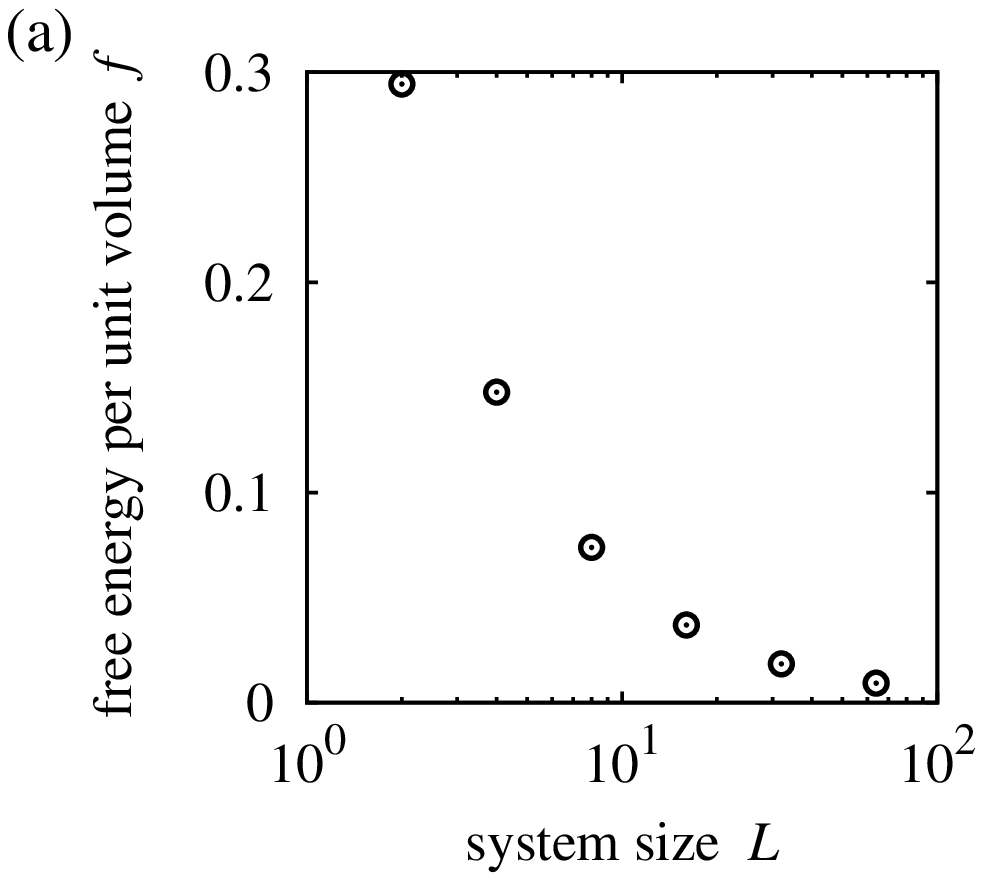}}
 {\includegraphics[width=0.95\linewidth,clip]{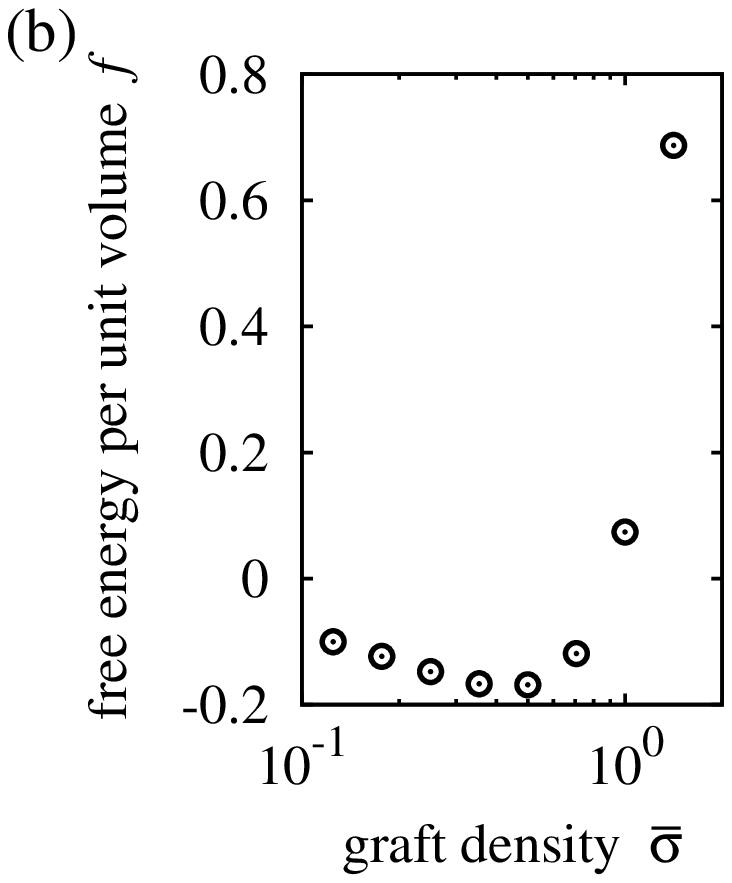}}
 \caption{}
 \label{free_energy_per_unit_volume}
\end{figure}

{
If the compressibility is not small, the density profile deviates from the
homogeneous profile. We show the simulations for compressible systems in
Appendix \ref{effect_of_compressiblity_on_density_and_potential_fields}.
In the cases of relatively large $\kappa$ values, the segment density
near the crystal-amorphous interface increases and the segment density
near the system center decreases. The strength of the potential layers
decreases as the density profile deviates from the homogeneous
one.

Before we analyze the tie and loop subchain statistics, here we briefly
examine the thermodynamic stability of the obtained structures.
The thermodynamic stability of the structure is determined by the free
energy per unit volume. Figure \ref{free_energy_per_unit_volume}
shows the dependence of the free energy per unit volume $f$ to the system
size $L$ and to the graft density $\bar{\sigma}$. From Figure
\ref{free_energy_per_unit_volume}(a), $f$ monotonically decreases
as $L$ increases. This means
that our system has no thermodynamically optimal size, as expected. On
the other hand, as observed in Figure
\ref{free_energy_per_unit_volume}(b), $f$ is not a monotonic function of
$\bar{\sigma}$. This means that there is the optimal graft density
which is thermally stable. However, we should recall that the graft density is not
freely tunable in our system, because it
is determined by the structure of the crystalline phase.
}

\subsection{Tie and Loop Distributions}
\label{tie_and_loop_distributions}

\begin{figure}[c!]
 \centering
 {\includegraphics[width=0.95\linewidth,clip]{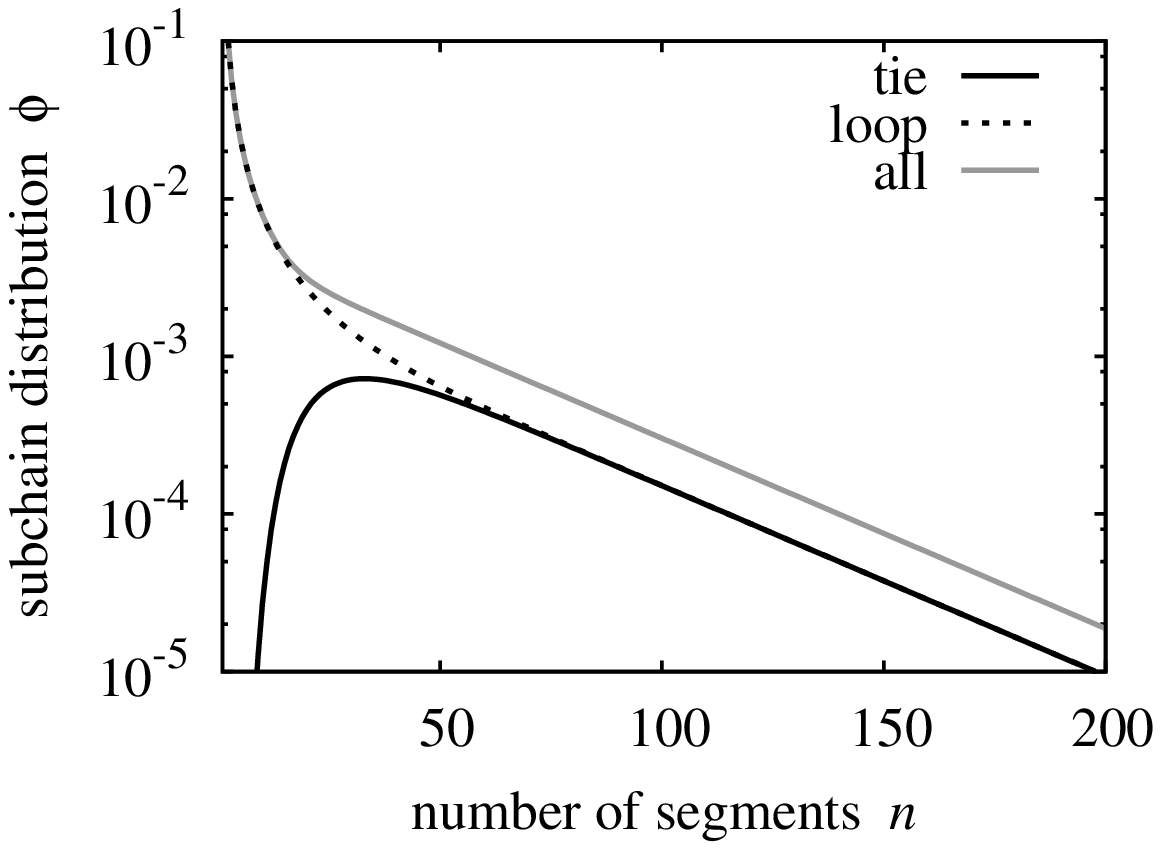}}
 \caption{}
 \label{subchain_fractions_s1_l8}
\end{figure}

Secondly, we show the tie and loop subchain distributions.
Figure \ref{subchain_fractions_s1_l8} shows the subchain distributions
for tie, loop, and all subchains.
The loop subchain distribution $\phi_{l}(n)$ decreases rapidly as $n$ increases.
This means that most of loop subchains are short. If we consider the
loop subchain with $n = 1$ as the tight loop which forms the adjacent
reentry\cite{Flory-1962}, the fraction of the tight
loop becomes $0.55$. This value is not so different from the prediction of
the gambler's ruin model\cite{Guttman-DiMarzio-1982}, $2 / 3$.
The average numbers of
segments for all subchains, $\bar{n}$, can be calculated as
\begin{equation}
 \label{average_number_of_segments}
 \bar{n} = \frac{\bar{\rho} L}{\bar{\sigma}} .
\end{equation}
Thus we have $\bar{n} = 8$ for the current system ($L = 8$, $\bar{\sigma} = 1$, and
$\bar{\rho} = 1$). The tie subchain distribution shown in Figure
\ref{subchain_fractions_s1_l8} has a maximum at $n = 33$, which is much
larger than $\bar{n}$. Thus most of the tie subchains consist of larger number
of segments than $\bar{n}$. For $n \gtrsim 50$, the tie and loop
subchain distributions almost coincide. As we mentioned, the probabilities that a
segment near the system center belongs to tie and loop subchains are
expected to be almost the same. Thus it seems to be natural that the tie
and loop subchains distributions almost coincide for large $n$.

\begin{figure}[c!]
 \centering
 {\includegraphics[width=0.95\linewidth,clip]{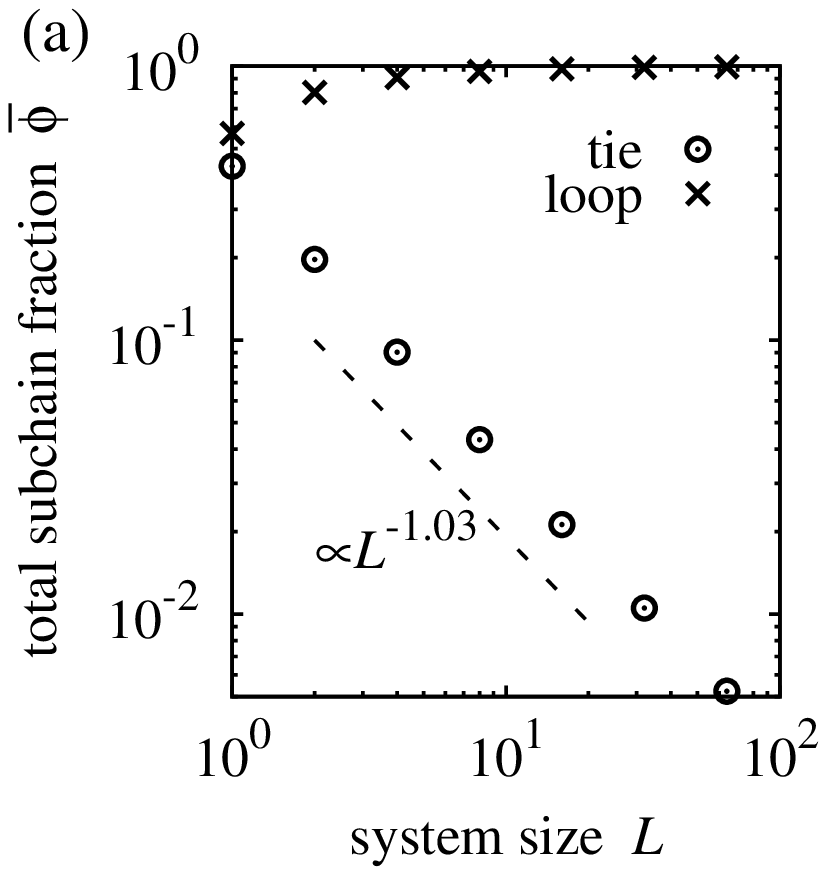}}
 {\includegraphics[width=0.95\linewidth,clip]{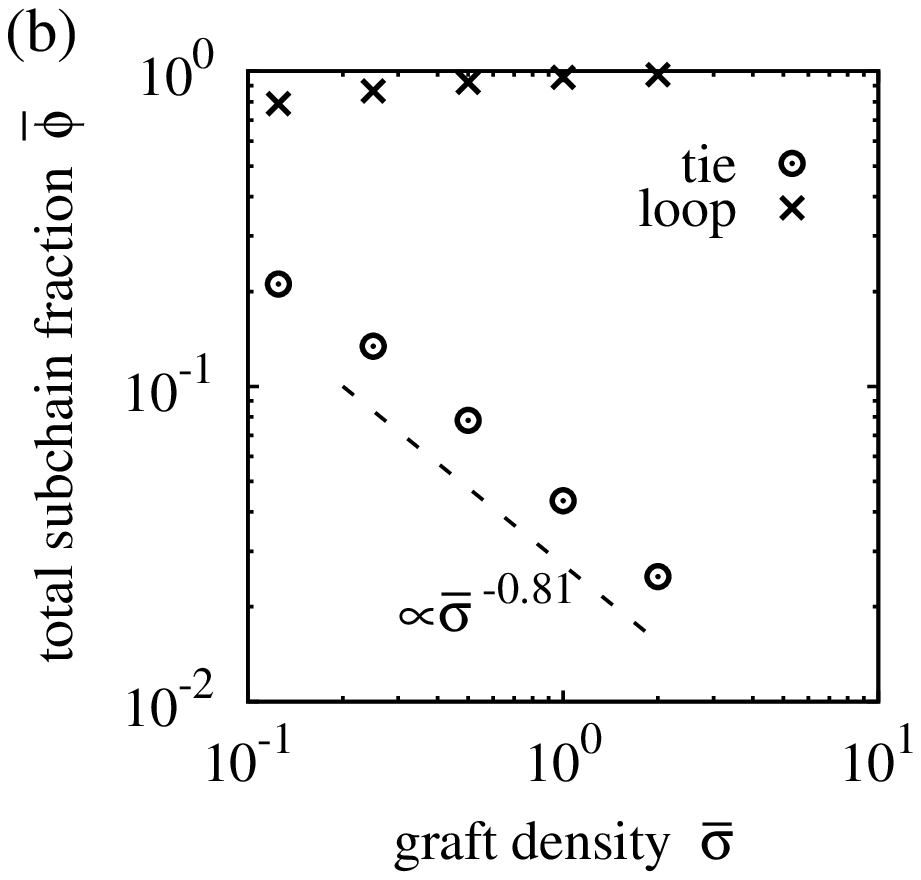}}
 \caption{}
 \label{total_subchain_fractions_dependence}
\end{figure}

We show the total tie and loop subchain fractions defined as
\begin{equation}
 \bar{\phi}_{l} = \sum_{n = 1}^{\infty} \phi_{l}(n), \qquad \bar{\phi}_{t} = \sum_{n = 1}^{\infty} \phi_{t}(n)
\end{equation}
for $\bar{\sigma} = 1$ in Figure \ref{total_subchain_fractions_dependence}(a). For large $L$, the total tie subchain fraction 
can be expressed by a simple power-law type relation $\phi_{t}(n)
\propto L^{-1.03}$.
We show the graft density dependence of the total tie and loop subchain fractions for $L = 8$ in Figure \ref{total_subchain_fractions_dependence}(b).
The $\bar{\sigma}$-dependence of $\bar{\phi}_{t}$ can also be fitted
well by a power-law. The fitting of the power-law to the
simulation data gives $\bar{\phi}_{t} \propto \bar{\sigma}^{-0.81}$.
{The decrease of the
total tie subchain fraction may be caused as a result of the increase of the total
loop subchain fraction by tight loop subchains.
As observed in Figure \ref{density_profiles_normalized}(b), the tie
segment density near the interface largely decreases as the graft
density increases. We expect that this corresponds the increase of tight
loops, and consequently the total tie subchain fraction decreases.
}

\begin{figure}[c!]
 \centering
 {\includegraphics[width=0.95\linewidth,clip]{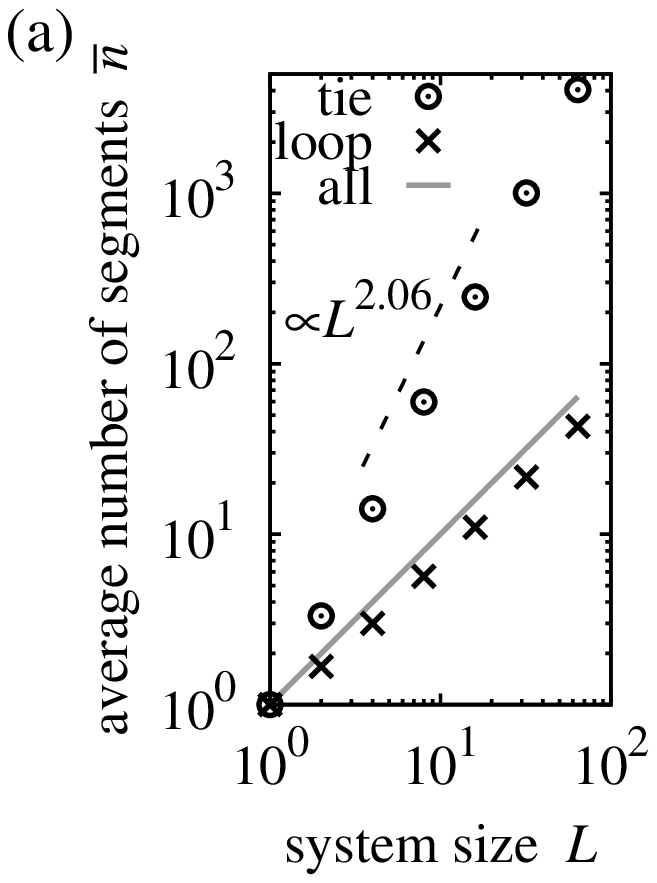}}
 {\includegraphics[width=0.95\linewidth,clip]{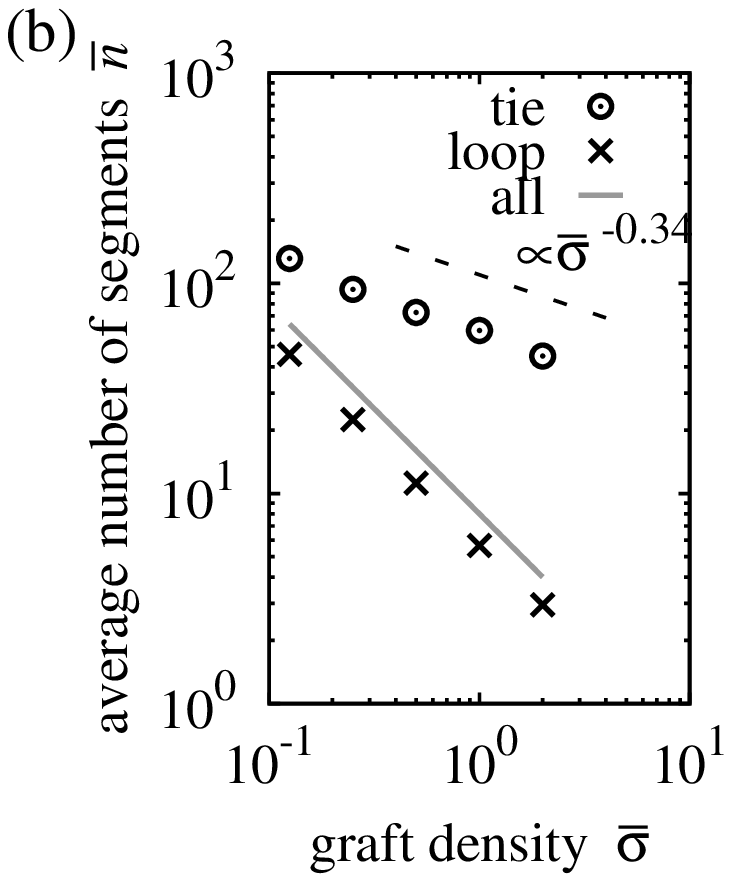}}
 \caption{}
 \label{average_numbers_of_segments_dependence}
\end{figure}

The average numbers of segments for tie and loop subchains,
$\bar{n}_{l}$ and $\bar{n}_{t}$, can be
calculated from the subchain distributions:
\begin{equation}
 \bar{n}_{l} = \frac{1}{\bar{\phi}_{l}} \sum_{n = 1}^{\infty} n \phi_{l}(n), \qquad
  \bar{n}_{t} = \frac{1}{\bar{\phi}_{t}}\sum_{n = 1}^{\infty} n
  \phi_{t}(n) .
\end{equation}
We show the system size dependence of the average numbers of segments
for tie, loop and all subchains in Figure
\ref{average_numbers_of_segments_dependence}(a).
The average numbers of segments for all subchains, $\bar{n}$, in Figure
\ref{average_numbers_of_segments_dependence}(a) is calculated by eq \eqref{average_number_of_segments}.
We also show the graft density dependence of the average numbers of
segments in Figure \ref{average_numbers_of_segments_dependence}(b).
For all the parameters examined above, $\bar{n}_{t}$ is always larger
than $\bar{n}_{l}$ and $\bar{n}$.
Although the total number fraction of the tie
subchain $\bar{\phi}_{t}$ is small, The tie subchains consist of large number of segments
compared with loop subchains.
We can observe that for large $L$, $\bar{n}_{t}$
depends on $L$ as $\bar{n}_{t} \propto L^{2.06}$. This $L$-dependence is
qualitatively different from ones of $\bar{n}$ and $\bar{n}_{l}$ 
($\bar{n}, \bar{n}_{l}
\propto L^{1}$). The $\bar{\sigma}$-dependence of
$\bar{n}_{l}$ is rather simple, $\bar{n}_{l} \propto
\bar{\sigma}^{-1}$.
The $\bar{\sigma}$-dependence of $\bar{n}_{t}$, $\bar{n}_{t}
\propto \bar{\sigma}^{-0.34}$ is again different from the cases of
$\bar{n}$ and $\bar{n}_{l}$.

\section{Discussions}
\label{discussions}

\subsection{Effective Potential Field}
\label{effective_potential_field}

As shown in Section \ref{density_and_potential_fields}, the effective
mean field potential is almost constant except the narrow regions near the
crystal-amorphous interfaces. A potential layer near an
crystal-amorphous interface consists of two sublayers. The first sublayer is
very thin and attractive, whereas the second sublayer is rather broad and repulsive.
This result is apparently inconsistent with the Milner's simulation
results\cite{Milner-2011}. In this subsection, we discuss the
repulsive layer in detail.

\begin{figure}[h!]
 \centering
 {\includegraphics[width=0.95\linewidth,clip]{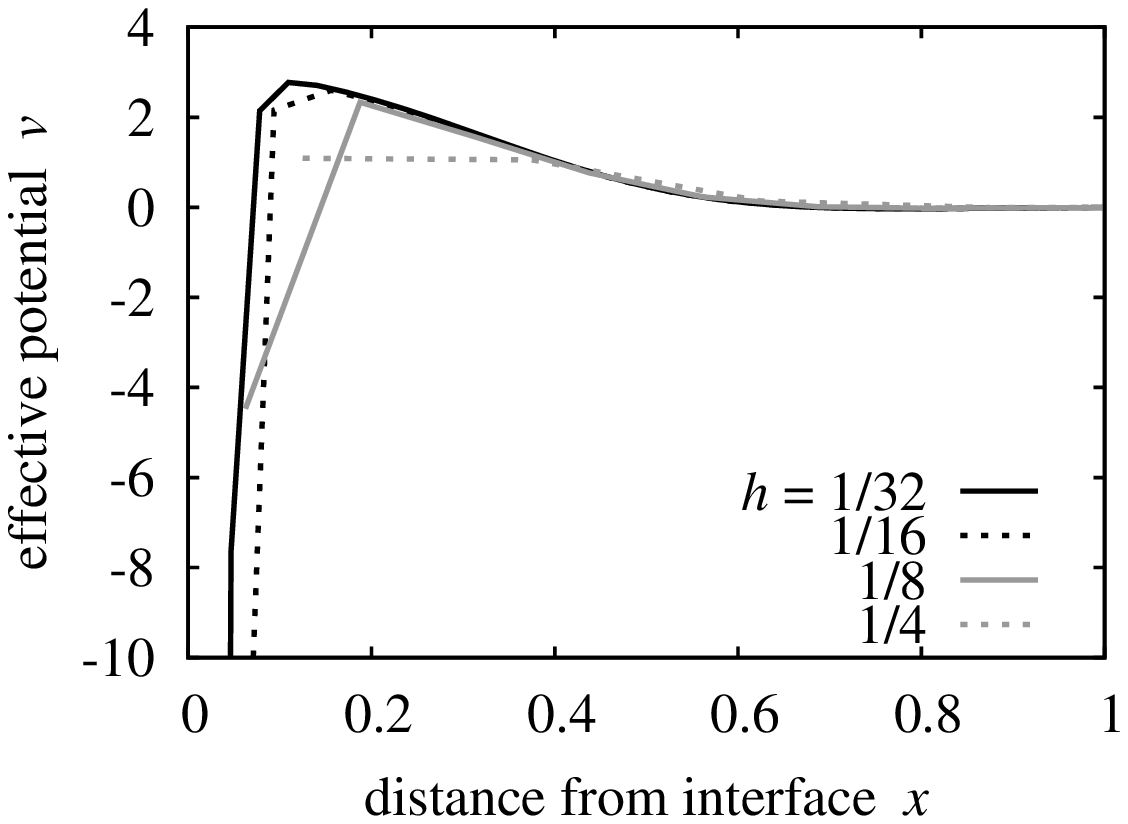}}
 \caption{}
 \label{potential_profile_s1_l8_mesh_size}
\end{figure}

\begin{figure}[h!]
 \centering
 {\includegraphics[width=0.95\linewidth,clip]{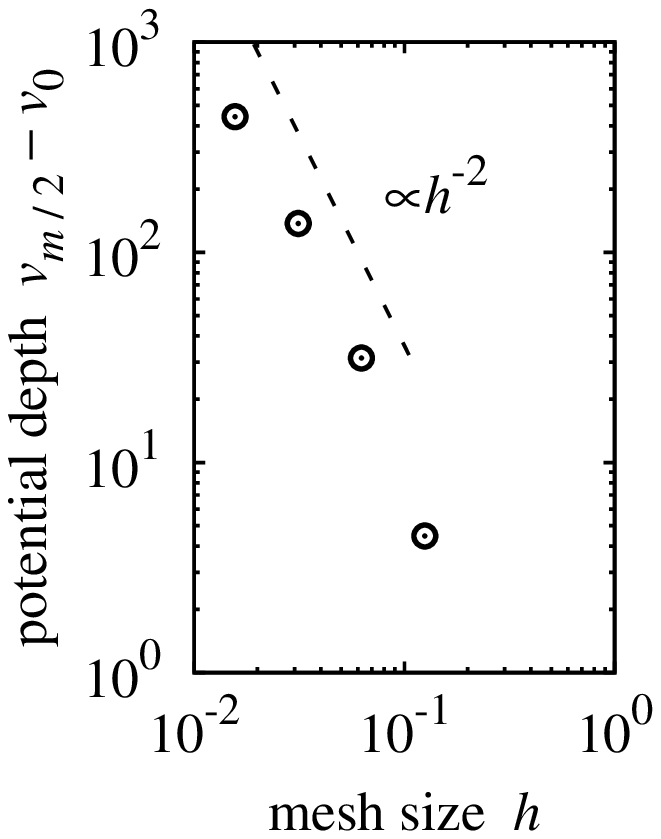}}
 \caption{}
 \label{interfacial_potential_s1_l8_mesh_size}
\end{figure}

The first sublayer is very thin and seems to be singular.
Whether this sublayer is singular or not can be easily confirmed by performing simulations
with different mesh sizes. Figure \ref{potential_profile_s1_l8_mesh_size}
shows the mean field potential
profiles calculated with different mesh sizes ($h = 1/32, 1/16, 1/8$ and
$1/4$) for $L = 8$ and $\bar{\sigma} = 1$. The potential near the wall depends on $h$ strongly, which
corresponds to the singular behavior. To examine the properly of the singular
layer, we show the mesh size dependence of the potential depth at the interface,
$v_{m / 2} - v_{0}$, in Figure \ref{interfacial_potential_s1_l8_mesh_size}. The potential depth dependence of the mesh
size is roughly estimated as $(v_{m / 2} - v_{0}) \propto h^{-2}$. This
mesh size dependence
is much stronger than one of the delta function ($\propto h^{-1}$), and thus the singular
potential layer cannot be interpreted as a delta function type
singular potential. {The interpretation of such
singularity is not trivial. As a possible explanation, we consider the
modulation of the boundary condition.}
Such a mesh size dependence can be realized if we
{phenomenologically} introduce the non-absorbing boundary condition\cite{deGennes-1969},
\begin{equation}
 \label{non_absorbing_boundary_condition_edwards_equation}
 \left. \frac{\partial q(x,s)}{\partial x} \right|_{x = 0}
  = - k q(0,s),
  \qquad  \left. \frac{\partial q(x,s)}{\partial x}
    \right|_{x = L}
  = k q(L,s) 
\end{equation}
instead of the absorbing boundary condition (eq \eqref{boundary_condition_edwards_equation}).
Here $k$ is a constant which represents the strength of the microscopic
potential near the interface.
{Generally, eq
\eqref{non_absorbing_boundary_condition_edwards_equation} reflects the
effect of the microscopic potential on the walls. In our system, we do
not have the microscopic potential on the interfaces and thus the origin
of eq \eqref{non_absorbing_boundary_condition_edwards_equation} is not
clear. It may arise from the competition between two strong constraints;
the initial condition \eqref{initial_condition_edwards_equation} and the
original boundary condition
\eqref{boundary_condition_edwards_equation}. (The initial condition tends
to increase the density near the interface whereas the original
boundary condition tends to reduce it.)}
If we discretize eq
\eqref{non_absorbing_boundary_condition_edwards_equation}, we have the
same mesh size dependence of the potential (as shown in Appendix \ref{discretization_of_non_absorbing_boundary_condition}).
The singular potential layer is smeared out
if we use relatively large mesh size ($h = 1/4$ in Figure
\ref{potential_profile_s1_l8_mesh_size}). The lattice SCF model roughly
corresponds to the case of $h \approx 1$, and thus we expect that the singular
layer cannot be observed in the lattice SCF simulations.
{Also, the lattice in the Milner's model does not deform,
which constrain the local and strong stretching of chains.}
We consider these are the
reason why our and Milner's models give apparently inconsistent
potential fields.

It would be worth mentioning here that such a boundary condition
makes numerical calculations difficult. The numerical constant $k$ in eq
\eqref{non_absorbing_boundary_condition_edwards_equation} should be determined
self-consistently, just as the potential field. Namely, the boundary
condition changes during the iterations. This makes the use of
the Fourier spectral method\cite{Matsen-chapter} difficult. Besides, the
update of the potential field can be large and unstable
if the update parameter $\theta$ is not small.
This is why the values of $\theta$ in our
simulations are much smaller than the usual SCF simulations.

The second sublayer is rather broad and gradually decays as the distance
from the interface increases. As shown in Figure \ref{potential_profile_s1_l8_mesh_size}, the profile of the
second sublayer is almost independent of the mesh size, and thus is not singular.
The profile of the second sublayer depends on $\bar{\sigma}$ rather strongly,
while it is almost independent of $L$. In the examined parameter range,
the width of the second sublayer is of the order of the segment size $b$
(which was taken to be unity in the simulations). This is consistent
with Milner's results.

From these observations, we consider that the
incompressible condition modulates the boundary condition and makes
the repulsive potential layers at the crystal-amorphous interfaces.
The importance of the boundary condition depends on the level of the
description. At fine levels (such as our simulations), the non-absorbing
boundary condition (or the singular potential layer) becomes
important. On the other hand,
at coarse levels (such as the lattice SCF simulations), the effect of the boundary condition (or
the singular sublayer) becomes not important and we may express the
effects of the incompressible condition only by thin repulsive potential
layers. {The subchain
statistics may be modeled in a similar way to the adsorption of polymer
chains onto surfaces\cite{deGennes-1976,Semenov-BonetAvalos-Johner-Joanny-1996}.
Actually, Milner\cite{Milner-2011} proposed to model a subchain as a
sequence of the trains and free loops. A similar model with two potential
layers may be constructed. In such a coarse-grained model, the potential
layers affect the statistics of the trains and free loops.
}

{We expect that the thickness of the repulsive potential
layer is determined as a result of the competition between the
incompressible and graft conditions. The strength of the
graft effect may be characterized by the graft density $\bar{\sigma}$.
We expect $\bar{\rho}$
can be utilized to characterize this condition (although the treatment
of the incompressible condition is not that simple).
Then, as a rough estimate, we expect that the ratio of these two parameters $\bar{\sigma} /
\bar{\rho}$ (which has the dimension of the length) determines the thickness of the repulsive layer. This
rough argument seems to be consistent with the results in Figure
\ref{potential_profile_half}. (The thickness increases as $\sigma$
increases, and is independent of $L$.)}

The existence of the non-absorbing boundary condition and the repulsive
potential layer (or the existence of the two potential layers) affects the statistics of
subchains in the amorphous region. The statistics of chains deviates
from the ideal Gaussian chain statistics without any
potential fields. This means that the naive assumption of the ideal
Gaussian chain statistics may lead qualitatively incorrect results
for some statistical quantities. For example, our simulation
results show that the chain statistics strongly depends on the graft density
$\bar{\sigma}$. However, some tie chain models are based on the
statistics which is independent of the graft density does not show such
graft density dependence.
The Huang-Brown type model\cite{Huang-Borwn-1999} assumes
that the tie chain statistics is essentially determined by the Gaussian
chain statistics in the melt state. In such models, the effect of the
graft density is not taken into account (at least explicitly).

{It is worth mentioning that the effective potential fields
obtained by our simulations (and by Milner's simulations) are
qualitatively much different from the potential fields in polymer brush
systems\cite{Milner-Witten-Cates-1988,Kawakatsu-book,Matsen-chapter}. In
the case of the melt brush, the density profile is the constant and one
chain end is grafted onto the wall. This situation is similar to our
system. However, the potential field in melt brush depends strongly on the
distance from the wall. This behavior is qualitatively different from
our result. This may be due to the difference in conditions for the
number of segments in a (sub)chain. In our model, the number of segments
in a subchain is not constant and subchains can adjust its length
freely. (Due to this mechanism, the constraints by the graft and
incompressibility may be very weak, except near the crystal-amorphous
interfaces.)
On the other hand, in a melt brush, the number of segments in a
grafted chain is constant. 
}

\subsection{Statistics of Tie Subchain}
\label{statistics_of_tie_subchain}

The statistics of the tie subchains deviates from one expected
from the ideal Gaussian chain statistics.
This would be observed clearly if we compare our simulation results with
the estimates based on the ideal Gaussian chain statistics without any constraints.
As shown in Appendix \ref{statistics_of_tie_subchain_without_constraints},
the tie subchain
fraction decays exponentially as $L$ increases {for large $L$}. This $L$-dependence is
qualitatively different from the SCF simulation result where the absorbing
boundary condition and the
incompressible condition are imposed. Namely, we underestimates the tie
subchain fraction if we assume the ideal Gaussian chain statistics.
Besides, without constraints, the average number of segments of a tie subchain depends $L$ as
$\bar{n}_{t} \propto L^{1.5}$ {(for large $L$)}, which is again qualitatively different
from the SCF simulation result. Thus, the statistics of ideal Gaussian
chains without any constraints cannot explain the statistics of
subchains between crystalline lamellae.
The boundary condition and
the interaction between segments are essential for the tie subchain statistics.

From the SCF simulation results, we expect that the tie subchain fraction
and the average number of segments of a tie subchain depend on the system size as
\begin{equation}
 \label{average_tie_fraction_and_segment_system_size_dependence}
 \bar{\phi}_{t} \propto L^{-1}, \qquad \bar{n}_{t} \propto L^{2} .
\end{equation}
Interestingly, this system size dependence of $\bar{n}_{t}$ is the same
as one of an ideal Gaussian chain. Although a tie 
subchain does not behave as an ideal Gaussian chain, some properties of the
tie subchain may be reasonably modelled with the ideal Gaussian chain statistics.
It should be mentioned here that these results are consistent with the
lattice-based gambler's ruin model\cite{Guttman-DiMarzio-1982}.
We consider that this is because the incompressible and impenetrable
conditions are satisfied in the gambler's ruin model.
Although the gambler's ruin model does not
utilize the effective mean field potential field, the incompressible
condition can control the conformation of subchains and thus the
resulting statistics becomes similar to the SCF model.
It would be worth mentioning that the SCF model presented in this work
is continuum and thus it is free from the limitations and artifacts of lattice-based
models. For example, our model can continuously change the parameters
such as $L$ and $\bar{\sigma}$.

\begin{figure}[c!]
 \centering
 {\includegraphics[width=0.95\linewidth,clip]{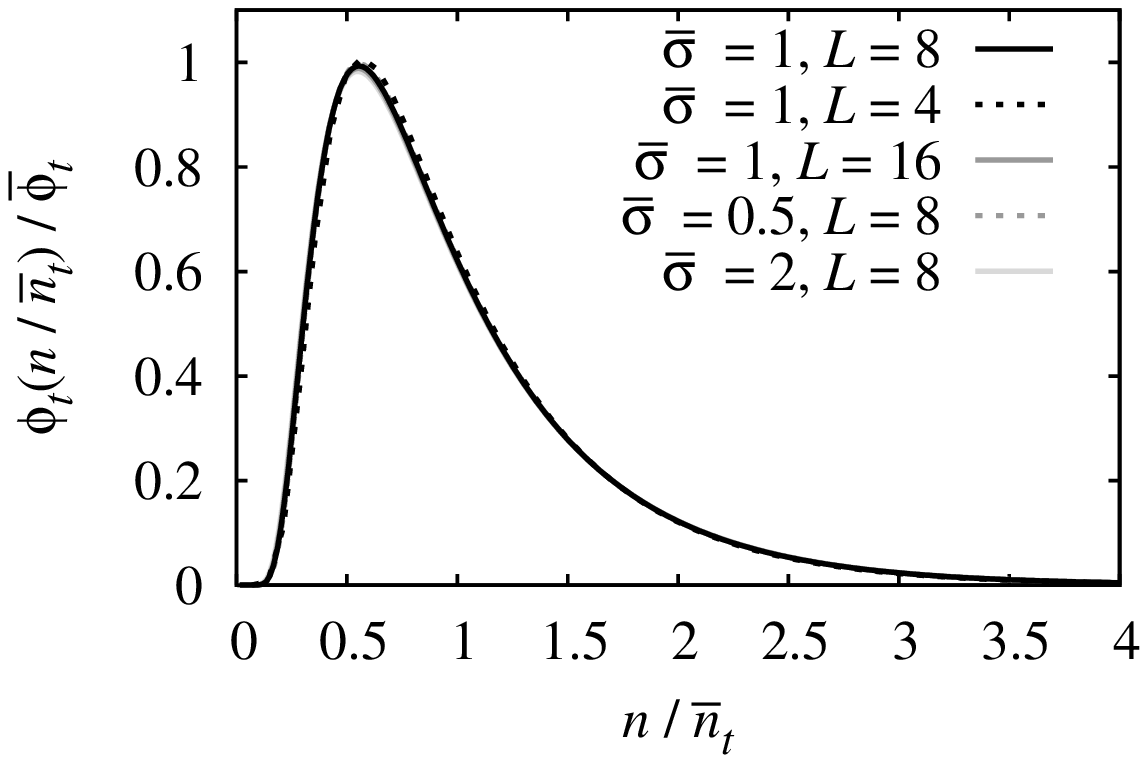}}
 \caption{}
 \label{master_curve_tie_distribution}
\end{figure}

From the result in Section \ref{effective_potential_field} together with the
simple system size relation (eq
\eqref{average_tie_fraction_and_segment_system_size_dependence}), we
expect that there may be a universality for the tie subchain statistics.
If the tie chain distribution is universal, we may be able to construct a master
curve for the tie chain distribution function. To examine whether such a
master curve can be constructed or not,
we rescale the number of segments $n$ and the tie chain distribution function
$\phi_{t}$ by the average number of segments $\bar{n}_{t}$ and the
total tie chain fraction $\bar{\phi}_{t}$, respectively. Figure
\ref{master_curve_tie_distribution} shows the rescaled tie
chain distribution functions for various different values of $L$ and
$\bar{\sigma}$. We can observe that all the rescaled tie chain
distributions collapse into one master curve within an acceptable error. Thus we conclude that the
tie chain distribution function is universal (at least in the examined
parameter range). Moreover, the master
curve can be reasonably fitted to the following empirical form:
\begin{equation}
 \label{fitting_function_master_curve_tie_distribtuion}
 \frac{\phi_{t}(n / \bar{n}_{t})}{\bar{\phi}_{t}}
  \propto \left(\frac{\bar{n}_{t}}{n}\right)^{2}
  \exp\bigg( - \alpha \frac{\bar{n}_{t}}{n} - \beta \frac{n}{\bar{n}_{t}} \bigg)
\end{equation}
where $\alpha$ and $\beta$ are fitting parameters. The fitting with the
SCF simulation data (shown in Figure \ref{master_curve_tie_distribution}) gives
$\alpha \approx 1.4$ and $\beta \approx 0.9$.

Here it would be fair to mention the limitaions of our SCF model.
Although our model gives interesting results on the statistics of
loop and tie subchains, it should be noticed that our model is based on
some assumptions. We should be careful whether the assumptions are
reasonable for the target systems or not. For example, if the molecular weight is
small, the effect of free ends (cilia) is not negligible.
The annealing may not fully equilibrate the subchains due to the very slow
relaxation by trapped entanglements (true entanglements).

{ The trapped entanglements would largely affect the chain
statistics. As an example, we consider two loop subchains which are
connected to different interfaces. If two loop subchains are entangled,
there are constraint on the conformations of subchains. (Unentangled
conformations are not allowed.) This constraint affects the path integral field and the
partition function. Intuitively, we expect that the long subchains are
easily entangled, and thus the constraint by trapped entanglements becomes
important especially for short subchains.
Unfortunately, as far as the
authors know, it is quite difficult to take into account the trapped
entanglements into the continuum field models such as the SCF model.
The extension of the SCF model, or the combination of the SCF and
molecular models \cite{Terzis-Theodorou-Stroeks-2000,Terzis-Theodorou-Stroeks-2000a,Daoulas-Muller-2006} would be required to study
the effect of trapped entanglements.
Despite these limitations, we believe that our model
can still provide non-trivial results, and can be utilized to study
statistics and properties of amorphous subchains.}

\subsection{Possible Applications}
\label{possible_applications}

Our simulation models can handle various parameters and is suitable
to determine some statistical quantities of the tie subchains, as shown
in Section \ref{results}. Here we discuss how our model and results can
be used for applications. A simple but interesting application is the 
combination of our simulation model with other coarse-grained simulation
models. The combined simulations will enable us to perform large scale simulations with small
computational costs. The modelling and simulations for larger systems (which
consist of multiple crystalline and amorphous layers) are left for
future works.

{The free energy per unit volume may be
utilized to analyze the multiple crystalline and amorphous layers.
As shown in Figure \ref{free_energy_per_unit_volume}(a), the free energy
per unit volume decreases as the thickness of the amorphous layer
increases. Intuitively, this can be interpreted as a repulsive interaction
between crystal-amorphous interfaces which stabilize the crystalline
lamellar structure.}

The simulation results obtained in this work can also be used as inputs for
theoretical models. For example, we can use the tie subchain distribution to estimate
the whole chain conformation in multiple crystalline and amorphous
layers. Such calculation gives the tie molecule fraction in annealed and
well-equilibrated systems. Becasue the tie subchain statistics has
rather simple properties (such as eqs
\eqref{average_tie_fraction_and_segment_system_size_dependence} and
\eqref{fitting_function_master_curve_tie_distribtuion}), some
statistical quantities may be simple.
We expect that the resulting tie modelcule statistics will be applied to various systems
(as long as the chains are well-equilibrated), and the comparison of thus estimated tie
molecule statistics with experimental data is an interesting future work.

If we use the effective potential field as a mean-field potential for
molecular simulations, some dynamical properties of amorphous subchains
can be studied. Although the potential layer is thin, it may
modulate the dynamics of amorphous subchains. We naturally expects that the segments near the
graft points will behave in a qualitatively different way from the segments apart
from the graft points. Such a picture is consistent with the
three-phase models for crystalline
polymers\cite{Sedighiamiri-vanErp-Peters-Govaert-vanDommelen-2010,Gueguen-Ahzi-Makradi-Belouettar-2010}.
In the three-phase models, there is an intermediate layer between crystalline and amorphous layers. Our
model will be useful to estimate the properties of the intermediate layers.

Another interesting application is the annealing dynamics. The dynamic
SCF model\cite{Kawakatsu-book,Muller-Schmid-2005} allows us to simulate the relaxational
dynamics. By combining our model with the dynamic SCF model,
the simulations for the relaxation of subchain distributions will be
possible. Such simulations will give the time evolution of the tie
subchain distribution. If we start the simulation from a nonequilibrium
subchain distribution, the rearrangement of subchains will be observed.
NMR experiments\cite{SchmidtRohr-Spiess-1991,Hedesiu-Demco-Kleppinger-Buda-Blumich-Remerie-Litvinov-2007}
showed that the polymer chains diffuse between the crstal and amorphous
regions during the annealing, and fraction of tight and semi-rigid amorphous chains changes by
annealing processes. The comparison of the simulation data and the NMR
data will be interesting.

\section{Conclusions}
\label{conclusions}

We proposed the SCF model for the amorphous chain statistics between
crystalline lamellae. We also proposed a numerical scheme for the SCF
model and preformed simulations with various different system sizes and
graft densities. In our model, the amorphous-crystal interfaces are
modelled as impenetrable surfaces, and a subchain is modelled as a graft chain.
We explicitly took into account the incompressible
condition, which originally arose from the interaction between segments.
The SCF simulation results showed that there exist
thin potential layers near the crystal-amorphous interfaces. A potential
layer consists of two sublayers. The first sublayer is a singular
attractive layer, which can be interpreted as the non-absorbing boundary
condition. The second sublayer is a relatively broad repulsive layer.
The statistics of subchains is affected by the potential layers.

The tie and loop subchain statistics were also calculated by the SCF
simulations. The tie subchains consist of larger number of segments than
the average. The system size dependence of
the average number of segments in the tie subchain and tie subchain
fraction obey rather simple statistics, and they are consistent with the
prediction of the gambler's ruin model.
We found that the tie subchain distribution function can be collapse
into one master curve, if we rescale the tie subchain distribution and
the segment number by the total tie subchain fraction and the average
number of segments in a tie subchain, respectively.

\section*{Acknowledgment}

This work was supported by Grant-in-Aid (KAKENHI) for Young
Scientists B 25800235. T.~U. thanks Dr.~Quan Chen (Penn State University)
for informing him about Ref \cite{DiMarzio-1965}.

\appendix
\section*{Appendix}

\section{Derivation of the Self-Consistent Field Equations}
\label{derivation_of_the_self_consistent_field_equations}

Although the SCF model itself has been studied widely and its derivation
for bulk systems is well known, the SCF model for grafted polymer
systems with a variable segment number is not trivial.
In this appendix, we show the detailed derivation of the SCF model in the main
text. We start from a multichain system in a three dimensional box. We express the number of subchains
of which segment number is $n$ as $M_{n}$, and the conformation of the
$j$-th subchain which consists of $n$ segments as $\bm{R}_{n,j}$.
From the nature of the crystalline structure, the segment number $n$
should be discrete. For simplicity, we assume that $n$ is an integer.
The partition function $\mathcal{Z}$ of the system is expressed as
\begin{equation}
 \label{partition_function_multichain}
 \mathcal{Z}
  = \sideset{}{'} \sum_{\lbrace M_{n} \rbrace}
  \bigg[ \prod_{n = 1}^{\infty} \frac{1}{M_{n}!} \bigg]
  \int \bigg[ \prod_{n = 1}^{\infty} \prod_{j = 1}^{M_{n}}
  \tilde{\mathcal{D}}\bm{R}_{n,j} \bigg] 
  \exp\bigg[ - \frac{\mathcal{U}[\hat{\rho}]}{k_{B} T}
      + \frac{\mu}{k_{B} T} \sum_{n = 1}^{\infty} n M_{n} \bigg] .
\end{equation}
Here, $\mu$ is the chemical potential for segments.
The summation over $\lbrace M_{n} \rbrace$ in
\eqref{partition_function_multichain} represents the summation
under the constraint
\begin{equation}
 \label{subchain_number_constraint}
 \sum_{n = 1}^{\infty} M_{n} = A \bar{\sigma} ,
\end{equation}
($A$ is the interfacial area) and $\int \tilde{\mathcal{D}}\bm{R}_{n,j}$ in eq
\eqref{partition_function_multichain} represents the functional integral
over the conformation of an end-grafted subchain with the statistical weight of the ideal Gaussian chain:
\begin{equation}
 \label{functional_integral_grafted_chain}
 \begin{split}
 \tilde{\mathcal{D}}\bm{R}_{n,j}
  & \equiv \mathcal{D}\bm{R}_{n,j}
   \, \exp\bigg[ - \int_{0}^{n} ds \,
  \frac{3}{2 b^{2}}
  \left|\frac{\partial \bm{R}_{n,j}(s)}{\partial s}\right|^{2} \bigg]\\
  & \qquad \times
  b^{2} \left[ \delta(R_{n,j,x}(0) - \epsilon) + \delta(R_{n,j,x}(0) - L + \epsilon)
  \right] \\
  & \qquad \times \left[ \delta(R_{n,j,x}(n) - \epsilon) + \delta(R_{n,j,x}(n)
  - L + \epsilon) \right] .
 \end{split} 
\end{equation}
Because the amorphous chain cannot penetrate into the crystalline
regions, the chain conformation should satisfy $0 \le R_{n,j,x}(s) \le
L$. The functional integral over the conformation is calculated under this constraint.
For simplicity, we assume that the measure for the functional integral
is determined appropriately so that the functional integral becomes
dimensionless.
$\hat{\rho}(\bm{r})$ is the microscopic segment density
field defined as
\begin{equation}
 \label{density_operator}
 \hat{\rho}(\bm{r}) \equiv \sum_{n = 1}^{\infty} \sum_{j = 1}^{M_{n}}
  \int_{0}^{n} ds \, \delta(\bm{r} - \bm{R}_{n,j}(s)),
\end{equation}
and $\mathcal{U}[\hat{\rho}]$ is the interaction potential between
segments:
\begin{equation}
 \label{interaction_energy_multichain}
 \mathcal{U}[\hat{\rho}]
  = \int d \bm{r} \, \frac{1}{2 \kappa \bar{\rho}^{2}}
 [\hat{\rho}(\bm{r}) - \bar{\rho}]^{2}  .
\end{equation}
Here, $\bar{\rho}$ and $\kappa$ in eq \eqref{interaction_energy_multichain}
are the average segment density
and the compressibility.

Following the standard procedure\cite{Muller-Schmid-2005}, we rewrite the partition function
\eqref{partition_function_multichain} in terms of the coarse-grained
segment density field.
Namely, we utilize the identity for the delta functional and
introduce the coarse-grained density field $\rho(\bm{r})$, 
\begin{equation}
 \label{delta_functional_identity}
 \begin{split}
  1 
  & = \int \mathcal{D}\rho \, \delta[ \rho - \hat{\rho}] \\
  & = \int \mathcal{D}\rho \mathcal{D}w \,
  \exp\bigg[ i \int d\bm{r} \, w(\bm{r}) [\rho(\bm{r}) - \hat{\rho}(\bm{r})] \bigg]
 \end{split}
\end{equation}
with $w(\bm{r})$ being an auxiliary potential field.
By inserting eq \eqref{delta_functional_identity}
into the right hand side of eq \eqref{partition_function_multichain} and utilizing the
Stirling's formula, we have
\begin{equation}
 \label{partition_function_multichain_modified}
 \mathcal{Z}
  = \sideset{}{'} \sum_{\lbrace M_{n} \rbrace} \int \mathcal{D}\rho
  \mathcal{D}w \, \exp
  \bigg[ - \frac{\mathcal{U}[\rho]}{k_{B} T} + i \int d\bm{r} \, w(\bm{r}) \rho(\bm{r})
   + \sum_{n = 1}^{\infty}
  M_{n} \ln \frac{e \mathcal{Q}(n)}{M_{n}}
   \bigg]
\end{equation}
where $\mathcal{Q}(n)$ is the partition function for a single subchain which
consists of $n$ segments:
\begin{equation}
 \label{single_chain_partition_function_n}
 \mathcal{Q}(n) \equiv
  e^{\mu n / k_{B} T} \int \tilde{\mathcal{D}}\bm{R}_{n} \,
  \exp\bigg[ - i \int_{0}^{n} ds \,
       w(\bm{R}_{n}(s)) \bigg] .
\end{equation}
The approximate solution (the mean field solution) can be obtained by maximizing the integrand in the right
hand side of eq \eqref{partition_function_multichain} with respect to
$\lbrace M_{n} \rbrace$, $\rho(\bm{r})$, and $w(\bm{r})$ (the saddle
point approximation). We
have the following set of equations as the saddle point condition:
\begin{align}
 & \label{saddle_point_equation_m}
 0 = \ln \frac{\mathcal{Q}(n)}{M_{n}} + \nu , \\
 & \label{saddle_point_equation_rho}
 0 = - \frac{1}{k_{B} T \kappa \bar{\rho}^{2} } [\rho(\bm{r}) -
 \bar{\rho}] + i w(\bm{r}) , \\
 & \label{saddle_point_equation_w}
 0 = i \rho(\bm{r}) - i \sum_{n = 1}^{\infty}
 \frac{M_{n}  e^{\mu n / k_{B} T}}{\mathcal{Q}(n)}
 \int_{0}^{n} ds  \int \tilde{\mathcal{D}}\bm{R}_{n} \,
 \delta(\bm{r} - \bm{R}_{n}(s)) \exp\bigg[ - i \int_{0}^{n} ds \,
  w(\bm{R}_{n}(s)) \bigg] .
\end{align}
Here we have introduced the Lagrange multiplier $\nu$ for the constraint
\eqref{subchain_number_constraint}. (Physically, this Lagrange multiplier can be
understood as the chemical potential for subchains.) From eqs
\eqref{subchain_number_constraint} and \eqref{saddle_point_equation_m},
the Lagrange multiplier $\nu$ is determined to be
\begin{equation}
 \label{lagrange_multiplier_m}
 \nu = \ln \frac{A \bar{\sigma}}{\mathcal{Q}} .
\end{equation}
where $\mathcal{Q}$ is the single chain partition function, 
\begin{equation}
 \label{single_chain_partition_function_sum}
 \mathcal{Q} \equiv \sum_{n = 1}^{\infty} \mathcal{Q}(n) .
\end{equation}
Eqs \eqref{saddle_point_equation_m} and \eqref{lagrange_multiplier_m} give the
explicit expression for the subchain number $M_{n}$, 
\begin{equation}
 \label{subchain_number}
 M_{n} = \frac{A \bar{\sigma} \mathcal{Q}(n)}{\mathcal{Q}} .
\end{equation}

For convenience, we define the mean field potential $v(\bm{r})$ as
\begin{equation}
 \label{mean_field_potential_definition}
 v(\bm{r}) \equiv i w(\bm{r}) .
\end{equation}
Because $w(\bm{r})$ is pure
imaginary at the saddle point, $v(\bm{r})$ is real.
By substituting eqs \eqref{subchain_number} and \eqref{mean_field_potential_definition} into eqs \eqref{saddle_point_equation_rho} and
\eqref{saddle_point_equation_w}, we have the modified saddle-point equations as
\begin{align}
 & \label{saddle_point_equation_rho_modified}
 v(\bm{r}) = \frac{1}{k_{B} T \kappa \bar{\rho}^{2} } [\rho(\bm{r}) -
 \bar{\rho}] , \\
 & \label{saddle_point_equation_w_modified}
 \rho(\bm{r}) = \frac{A \bar{\sigma}}{\mathcal{Q}} \sum_{n = 1}^{\infty}
 e^{\mu n / k_{B} T}
 \int_{0}^{n} ds  \int \tilde{\mathcal{D}}\bm{R}_{n} \,
 \delta(\bm{r} - \bm{R}_{n}(s)) \exp\left[ - \int_{0}^{n} ds \,
  v(\bm{R}_{n}(s)) \right] .
\end{align}
{
The free energy of the system under the saddle point approximation becomes
\begin{equation}
 \label{free_energy_saddle_point}
   \frac{\mathcal{F}}{k_{B} T}
   = \ln \mathcal{Z}
   \approx \frac{\mathcal{U}[\rho]}{k_{B} T} - \int d\bm{r} \, v(\bm{r}) \rho(\bm{r})
   - A \bar{\sigma} \ln \frac{e \mathcal{Q}}{A \bar{\sigma}} .
\end{equation}
}
From the symmetry of the system, $v(\bm{r})$ and $\rho(\bm{r})$ do not depend on
$y$ nor $z$. This means that the system can be treated as one
dimensional. Therefore it is sufficient for us to consider the physical quantities per unit
interfacial area. This is equivalent to simply set $A = 1$. The one dimensional version of eq
\eqref{saddle_point_equation_rho_modified} is eq
\eqref{potential_from_density}. The one dimensional version of eq
\eqref{saddle_point_equation_w_modified} can be rewritten in terms of
the path integral field and the partial differential equation, by using
the Feynman-Kac formula\cite{Oksendal-book}. 
The resulting equations are 
eqs \eqref{edwards_equation}-\eqref{density_from_path_integral}.
Similarly, eqs \eqref{single_chain_partition_function_n} and
\eqref{single_chain_partition_function_sum} gives eq \eqref{single_chain_partition_function}.
In addition, we impose the condition that the total number of segments
in the system is constant, and regard the chemical potential $\mu$ as the Lagrange
multiplier for this condition. This gives eq
\eqref{chemical_potential_condition} as the constraint.
{The one dimensonal version of eq
\eqref{free_energy_saddle_point} with the harmonic potential is eq \eqref{free_energy}.}
Thus we have the SCF model
in the main text (the set of eqs \eqref{edwards_equation}-\eqref{free_energy}).

\begin{figure}[h!]
 \centering
 {\includegraphics[width=0.95\linewidth,clip]{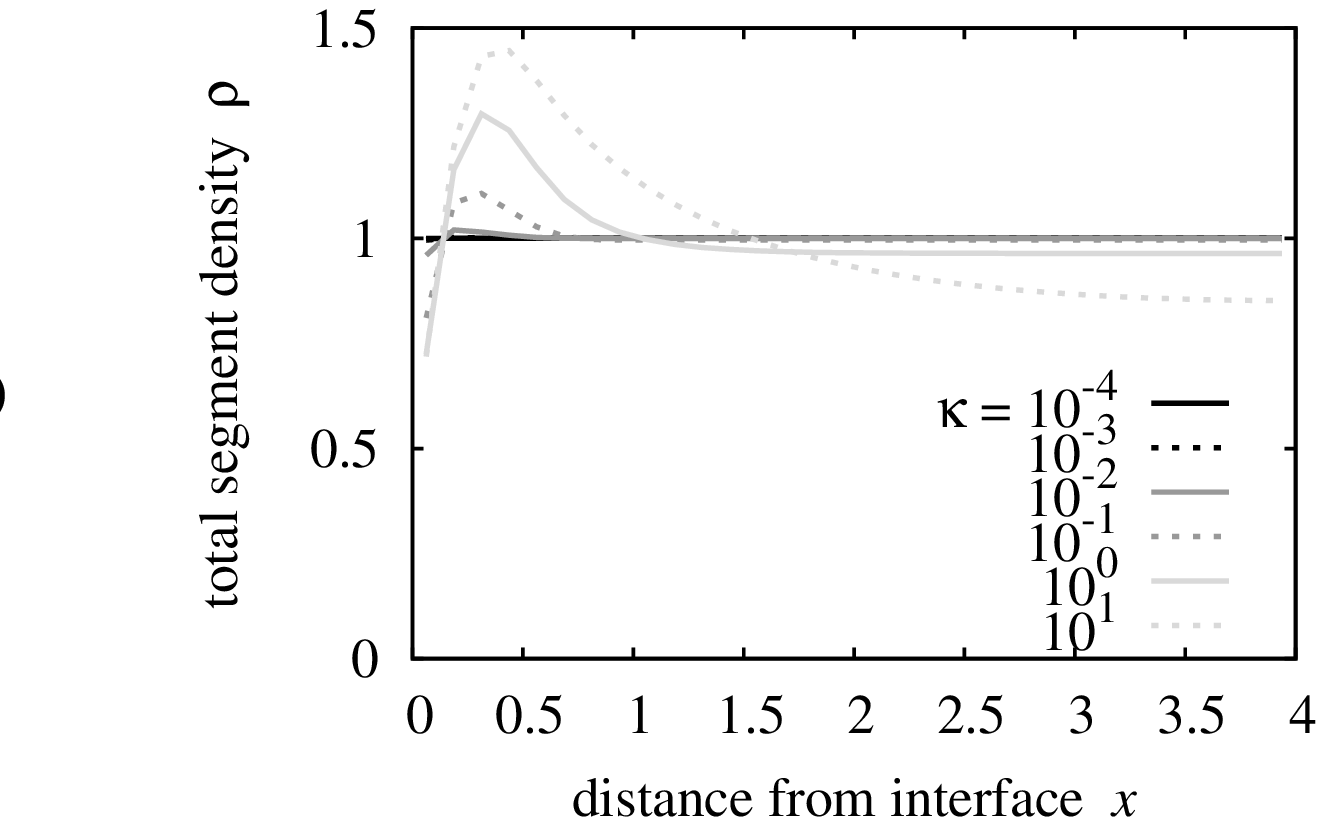}}
 \caption{}
 \label{density_profiles_s1_l8_half}
\end{figure}

\begin{figure}[h!]
 \centering
 {\includegraphics[width=0.95\linewidth,clip]{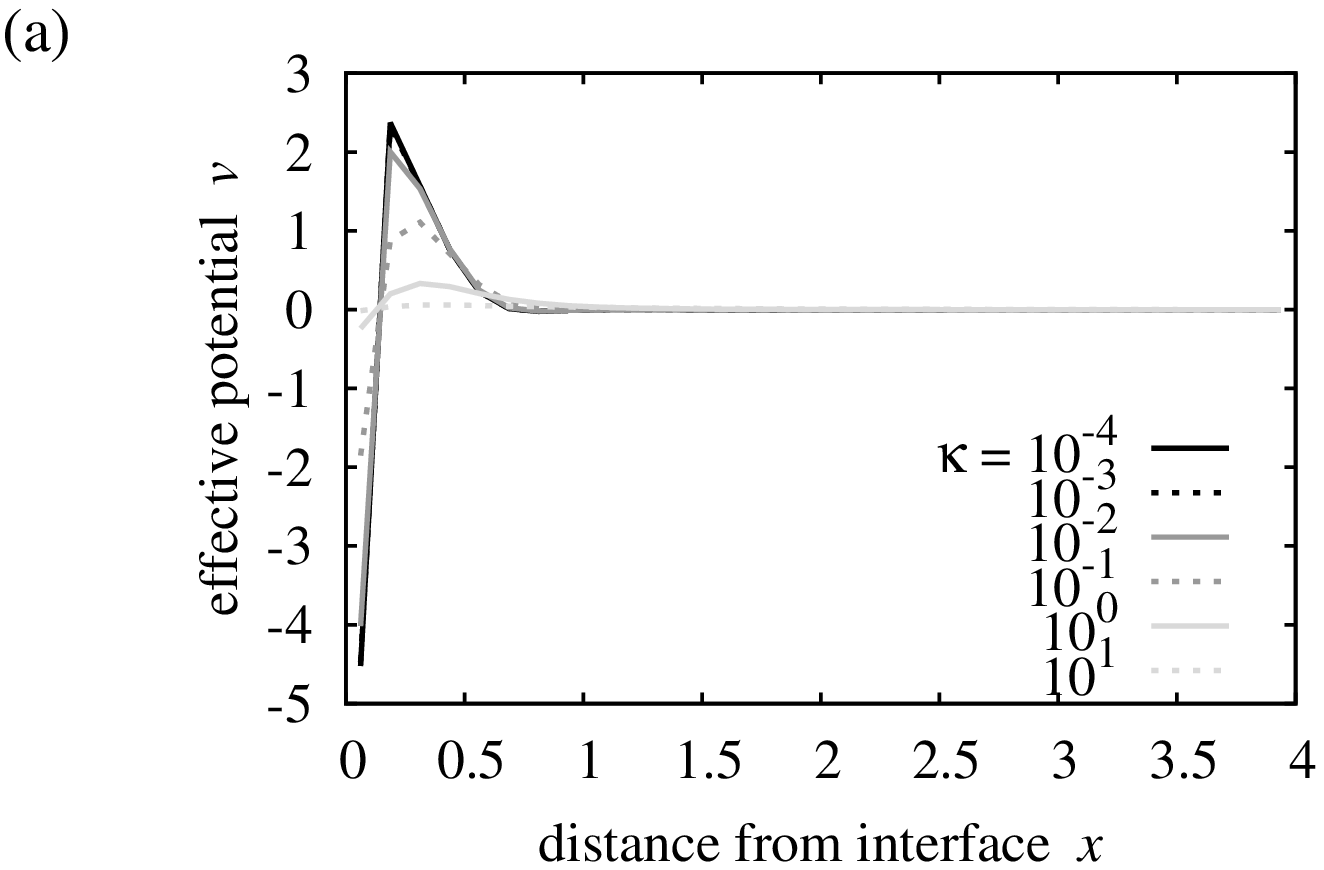}}
 \caption{}
 \label{potential_profile_s1_l8_half}
\end{figure}

{
\section{Effect of Compressibility on Density and Potential Fields}
\label{effect_of_compressiblity_on_density_and_potential_fields}
In the SCF simulations in the main text, we empirically choose $\kappa =
0.001$. As we mentioned, this value of $\kappa$ is sufficient to
reproduce almost incompressible density profiles. In this appendix, we
perform simulations with different $\kappa$ values ($\kappa = 10^{-4}
\sim 10^{1}$) and discuss what
happens if the system becomes compressible.

We show the total segment density field $\rho(x)$ and the effective
potential $v(x)$ for various values of $\kappa$ (and $\bar{\sigma} = 1$
and $L = 8$) in Figures
\ref{density_profiles_s1_l8_half} and \ref{potential_profile_s1_l8_half}.
We can observe that if $\kappa$ is sufficiently small ($\kappa =
10^{-3}$ and $10^{-4}$), both the density and potential profiles become
almost $\kappa$-independent. Therefore, 
we consider the system is practically incompressible for $\kappa \lesssim 10^{-3}$.
This justifies our empirical choice of $\kappa =
10^{-3}$ in the main text.

As $\kappa$ increases, the potential and density profiles deviate from
those for the incompressible system. From Figure
\ref{density_profiles_s1_l8_half}, we can observe that the density near
the wall increases and the density near the system center
decreases, as $\kappa$ increases. From Figure \ref{potential_profile_s1_l8_half}, we can also observe
the height of the repulsive potential layer decreases as $\kappa$ increases.
These results are consistent with our discussion in the main text.
The potential layer near the wall repel the segments to make the density
field constant. Conversely, for small
$\kappa$ cases, the potential layer does not work well and the segments
become relatively concentrated near the wall. The reduction of the
density near the system center can be directly related to the reduction
of the tie subchains, because we have the relation $\rho_{t}(L / 2) =
\rho(L / 2) / 2$ at the system center and the tie segments are
concentrated near the system center. As a result, the fraction of the
tie subchains decreases as $\kappa$ increases.
}

\section{Discretization of Non-Absorbing Boundary Condition}
\label{discretization_of_non_absorbing_boundary_condition}

In this appendix, we show that the discretization of the non-absorbing
gives singular potential
\eqref{non_absorbing_boundary_condition_edwards_equation} at the interface.
Eq \eqref{non_absorbing_boundary_condition_edwards_equation} can be
discretized as
\begin{equation}
 \label{non_absorbing_boundary_condition_edwards_equation_discretized}
  \frac{q_{0} - q_{-1}}{h}
  = - k \frac{q_{0} + q_{-1}}{2},
  \qquad  \frac{q_{m} - q_{m - 1}}{h}
  = k \frac{q_{m} + q_{m - 1}}{2}  .
\end{equation}
Here $q_{0}$ and $q_{m}$ are virtual path integral fields at the lattice
points at $x_{-1} = - h / 2$ and $x_{m} = L + h / 2$, respectively.
We can modify eq
\eqref{non_absorbing_boundary_condition_edwards_equation_discretized} as
follows, to obtain the explicit expressions for $q_{-1}$ and $q_{m}$.
\begin{equation}
 \label{non_absorbing_boundary_condition_edwards_equation_discretized_modified}
  q_{-1} = \frac{1 + k h / 2}{1 - k h / 2} q_{0}, \qquad
  q_{m} = \frac{1 + k h / 2}{1 - k h / 2} q_{m - 1} .
\end{equation}
The second order derivatives of $q(x)$ at $x_{0}$ and $x_{m - 1}$ are then
discretized as
\begin{align}
 & \label{second_order_derivative_discretized_0}
 \frac{q_{-1} - 2 q_{0} + q_{1}}{h^{2}} = 
   \frac{q_{-1} - 2 q_{0}}{h^{2}} + \frac{1 + k h /
   2}{h^{2} (1 - k h / 2)} q_{0} , \\
 & \label{second_order_derivative_discretized_m1}
 \frac{q_{m - 2} - 2 q_{m - 1} + q_{m}}{h^{2}} = 
   \frac{q_{m - 2} - 2 q_{m}}{h^{2}} + \frac{1 + k h /
   2}{h^{2} (1 - k h / 2)} q_{m} .
\end{align}
The first terms in the right hand sides of eqs
\eqref{second_order_derivative_discretized_0} and
\eqref{second_order_derivative_discretized_m1} correspond to the
discretized second order derivatives under the absorbing boundary
condition \eqref{boundary_condition_edwards_equation}.
The coefficient matrix $C_{ij}$ for the absorbing boundary
condition is recovered if we replace $v_{0}$ and
$v_{m - 1}$ in eq \eqref{coefficient_matrix} by
\begin{equation}
 v_{0}^{\text{(eff)}} = v_{0} - \frac{1 + k h /
   2}{6 h^{2} (1 - k h / 2)}, \qquad
 v_{m - 1}^{\text{(eff)}} = v_{m - 1} - \frac{1 + k h /
   2}{6 h^{2} (1 - k h / 2)} .
\end{equation}
(The potentials at
other lattice points ($v_{1}, v_{2}, \dotsb v_{m - 2}$) in eq
\eqref{coefficient_matrix} are not affected.)
$v_{0}^{\text{(eff)}}$ and $v_{m - 1}^{(\text{eff})}$ can be interpreted
as the effective potential field at the interface. If the mesh size is
small, they depend on $h$ as $v_{0}^{\text{eff}} \approx v_{m -
1}^{\text{(eff)}} \approx 1 / 6 h^{2}$. Thus we find that the
non-absorbing boundary condition corresponds singular potential layers and
the potential depth is proportional to $h^{-2}$. 

\section{Statistics of Tie Subchain without Constraints}
\label{statistics_of_tie_subchain_without_constraints}

In this appendix, we consider the systems without any
constraint. Namely, we do not impose the incompressible constraint nor
the absorbing boundary condition. 
The total tie subchain fraction can be estimated as follows.
From the ideal Gaussian chain statistics,
the probability
to find a loop or tie subchain is the same as the probability to
find the subchain of which end-to-end distance is $0$ or $L$, respectively.
Then, the tie and loop subchain distributions are approximately given as follows:
\begin{align}
 & \phi_{l}(n)
  \approx \frac{1}{\mathcal{Z} \sqrt{n}} e^{- \tilde{\mu} n} , \\
 & \phi_{t}(n)
  \approx \frac{1}{\mathcal{Z} \sqrt{n}} e^{- \tilde{\mu} n - 3 L^{2} /
 2 n b^{2}} .
\end{align}
Here we have defined $\tilde{\mu} \equiv - \mu / k_{B} T > 0$, and
$\mathcal{Z}$ is the partition function:
\begin{equation}
 \mathcal{Z} = \sum_{n = 1}^{\infty}
  \frac{1}{\sqrt{n}} e^{- \tilde{\mu} n}
  (1 + e^{- 3 L^{2} / 2 n b^{2}}) .
\end{equation}
{If $L$ is not very small,}
the sum over $n$ can be reasonably approximated by
the integral over $n$. Then we have
{
\begin{equation}
 \mathcal{Z}
  \approx \int_{0}^{\infty} dn \, \frac{1}{\sqrt{n}} e^{-\tilde{\mu} n}
  (1 + e^{-3 L^{2} / 2 n b^{2}}) =
  \sqrt{\frac{\pi}{\tilde{\mu}}}
  ( 1 + e^{-\sqrt{6 \tilde{\mu} L^{2} / b^{2}}} )
\end{equation}
}
and the average number of segments $\bar{n}$ is given as
{
\begin{equation}
 \label{average_number_of_segments_noninteracting}
 \bar{n} = - \frac{\partial \ln \mathcal{Z}}{\partial \tilde{\mu}}
  \approx \frac{1}{2 \tilde{\mu}}
  \bigg[ 1 +
  \frac{\sqrt{6 \tilde{\mu} L^{2} / b^{2}} }
  {e^{\sqrt{6 \tilde{\mu} L^{2} / b^{2}}} + 1} \bigg].
\end{equation}
}
Substituting eq \eqref{average_number_of_segments} into eq
\eqref{average_number_of_segments_noninteracting} 
gives {the relation between $\tilde{\mu}$ and $L$:
{
\begin{equation}
 \label{effective_chemical_potential_segments_noninteracting}
  \tilde{\mu} \approx \frac{ \bar{\sigma}}{2 \bar{\rho} L}
  \bigg[ 1 +
  \frac{\sqrt{6 \tilde{\mu} L^{2} / b^{2}} }
  {e^{\sqrt{6 \tilde{\mu} L^{2} / b^{2}}} + 1} \bigg]. 
\end{equation}
}
Although it is difficult to obtain the explicit expression of
$\tilde{\mu}$, we find that 
the second term in the parenthesis in the right hand side of eq
\eqref{effective_chemical_potential_segments_noninteracting} becomes
small both for small and large $L$ cases. Thus we simply approximate
$\tilde{\mu}$ as $\tilde{\mu} \approx \bar{\sigma} / 2 \bar{\rho} L$.
}
The total tie subchain fraction $\bar{\phi}_{t}$ is calculated as
{
\begin{equation}
 \label{total_tie_subchain_fraction_noninteracting}
  \bar{\phi}_{t} 
  \approx \frac{1}{\mathcal{Z}}
  \int_{0}^{\infty} dn \, \frac{1}{\sqrt{n}} e^{- \tilde{\mu} n - 3 L^{2} / 2 n b^{2}}
 \approx \frac{1}{e^{\sqrt{3 \bar{\sigma} L / \bar{\rho} b^{2}}} + 1} .
\end{equation}
}
The average number of segments of a tie subchain can be calculated in a similar
way:
\begin{equation}
 \label{average_segment_tie_subchain_noninteracting}
  \bar{n}_{t}
  \approx \frac{1}{\bar{\phi}_{t} \mathcal{Z}} \int_{0}^{\infty} dn \,
  n \frac{1}{\sqrt{n}}
  e^{- \tilde{\mu} n - 3 L^{2} / 2 n b^{2}}
  \approx \frac{\bar{\rho} L}{\bar{\sigma}}
  \bigg( 1 + \sqrt{\frac{3 \bar{\sigma} L}{\bar{\rho} b^{2}}} \bigg) .
\end{equation}
If $L$ is sufficiently large, eqs
\eqref{total_tie_subchain_fraction_noninteracting} and
\eqref{average_segment_tie_subchain_noninteracting} can simply reduces to
{
\begin{equation}
 \label{total_tie_subchain_fraction_average_segment_tie_subchain_noninteracting_approx}
  \bar{\phi}_{t}
 \approx e^{- \sqrt{3 \bar{\sigma} L / \bar{\rho}
 b^{2}}} , \quad  
 \bar{n}_{t}
 \approx
\sqrt{\frac{3 \bar{\rho} L^{3}}{\bar{\sigma} b^{2}}}.
\end{equation}
}
Therefore, for large $L$, the total tie subchain fraction decays exponentially as $L^{1/2}$
increases. The average number of segments of a tie subchain depends on $L$
as $\bar{n}_{t} \propto L^{3/2}$. As mentioned in the main text, these
$L$-dependence is qualitatively different from the SCF simulation results.


\clearpage

\section*{Figure Captions}

Figure \ref{lamellar_image}:
Schematic illustrations of (a) a crystalline lamellar structure and (b) an
amorphous region in the crystalline lamellae.
Grey and white colors represent crystal and amorphous regions,
respectively, and solid curves represent polymer chains. A tie subchain
connect two different crystalline lamellae whereas a loop subchain does not.

\vspace{\baselineskip}
\hspace{-\parindent}%
Figure \ref{density_profiles_s1_l8}:
The segment densities of tie, loop, and all subchains, for $L = 8$ and
$\bar{\sigma} = 1$. Two dotted curves represent the segment densities of
the loop subchain of which ends are grafted to $x = 0$ or $x = L$.

\vspace{\baselineskip}
\hspace{-\parindent}%
Figure \ref{density_profiles_normalized}:
The system size and graft density dependence of the tie segment density.
(a) The system size dependence for $\bar{\sigma} = 1$, and (b) the graft density dependence for $L = 8$.
The normalized position $x / L$ is used.

\vspace{\baselineskip}
\hspace{-\parindent}%
Figure \ref{potential_profile_s1_l8}:
The effective potential profile for segments for $L = 8$ and
$\bar{\sigma} = 1$. The potential is shifted so that the potential
becomes zero at the system center, $v(L / 2) = 0$.

\vspace{\baselineskip}
\hspace{-\parindent}%
Figure \ref{potential_profile_half}:
The system size and graft density dependence of the effective potential profile.
(a) The system size dependence for $\bar{\sigma} = 1$, and (b) the graft
density dependence for $L = 8$.
Only the profiles for $x \le L / 2$ is shown.

\vspace{\baselineskip}
\hspace{-\parindent}%
{
Figure \ref{free_energy_per_unit_volume}:
The system size and graft density dependence of the free energy per unit
volume. (a) The system size dependence for $\bar{\sigma} = 1$ and (b)
the graft density dependence for $L = 8$.}

\vspace{\baselineskip}
\hspace{-\parindent}%
Figure \ref{subchain_fractions_s1_l8}:
The subchain distributions for tie, loop, and all subchains. The system
size and graft density are $L = 8$ and $\bar{\sigma} = 1$.

\vspace{\baselineskip}
\hspace{-\parindent}%
Figure \ref{total_subchain_fractions_dependence}:
The system size and graft density dependence of the total tie and loop subchain fractions.
(a) The system size dependence for $\bar{\sigma} = 1$, and (b) the graft
density dependence for $L = 8$. The dashed lines represent the
power-law type relations $\bar{\phi}_{t} \propto L^{-1.03}$ and $\bar{\phi}_{t} \propto \bar{\sigma}^{-0.81}$.

\vspace{\baselineskip}
\hspace{-\parindent}%
Figure \ref{average_numbers_of_segments_dependence}:
The system size and graft density dependence of the average number of segments for tie,
loop, and all subchains.
(a) The system size dependence for  $\bar{\sigma} = 1$, and (b) the
graft density dependence for $L = 8$. The dashed lines represent the
power-law type relations $\bar{n}_{t} \propto L^{2.06}$ and $\bar{n}_{t} \propto \bar{\sigma}^{-0.34}$.

\vspace{\baselineskip}
\hspace{-\parindent}%
Figure \ref{potential_profile_s1_l8_mesh_size}:
The effective potential profile for $L = 8$ and
$\bar{\sigma} = 1$, calculated with various values of $h$. The strong
$h$-dependence of the potential
profile near $x = 0$ implies {the existence of a singular layer at $x =
0$}.

\vspace{\baselineskip}
\hspace{-\parindent}%
Figure \ref{interfacial_potential_s1_l8_mesh_size}:
The mesh size dependence of the potential depth at the crystal-amorphous
interface, $v_{m / 2} - v_{0}$. The dotted curve represents the
power-law type relation $v_{m / 2} - v_{0} \propto h^{-2}$.

\vspace{\baselineskip}
\hspace{-\parindent}%
Figure \ref{master_curve_tie_distribution}:
The rescaled tie subchain distribution functions for various $L$ and
$\bar{\sigma}$. The number of segments $n$ and the tie chain
distribution function $\phi_{t}$ are rescaled by the average
number of segment in a tie subchain $\bar{n}_{t}$ and the total tie
subchain fraction $\bar{\phi}_{t}$, respectively.

{
\vspace{\baselineskip}
\hspace{-\parindent}%
Figure \ref{density_profiles_s1_l8_half}:
The compressibility dependence of the total density for $\bar{\sigma} =
1$ and $L = 8$. 
Only the profiles for $x \le L / 2$ is shown.

\vspace{\baselineskip}
\hspace{-\parindent}%
Figure \ref{potential_profile_s1_l8_half}:
The compressibility dependence of the effective potential field for $\bar{\sigma} =
1$ and $L = 8$. 
Only the profiles for $x \le L / 2$ is shown.
}

\clearpage

\section*{Figures}

\clearpage

\clearpage

\clearpage

\clearpage

\clearpage

\clearpage

\clearpage

\clearpage

\clearpage

\clearpage

\clearpage

\end{document}